# Gamma- and Cosmic-Ray Observations with the GAMMA-400 Gamma-Ray Telescope

N.P. Topchiev[a,*], A.M. Galper[a,b], I.V. Arkhangelskaja[b], A.I. Arkhangelskiy[b], A.V. Bakaldin[c], R.A. Cherniy[a], I.V. Chernysheva[b], E.N. Gudkova[a], Yu.V. Gusakov[a], O.D. Dalkarov[a], A.E. Egorov[a], M.D. Kheymits[b], M.G. Korotkov[b], A.A. Leonov[a,b], A.G. Malinin[b], V.V. Mikhailov[b], A.V. Mikhailova[b], P.Yu. Minaev[a,d], N.Yu. Pappe[a], M.V. Razumeyko[a], M.F. Runtso[b], Yu.I. Stozhkov[a], S.I. Suchkov[a], Yu.T. Yurkin[b]

[a] Lebedev Physical Institute of the Russian Academy of Sciences, Moscow 119991, Russia
[b] National Research Nuclear University "MEPhI" (Moscow Engineering Physics Institute), Moscow 115409, Russia
[c] Scientific Research Institute for System Analysis of the Russian Academy of Sciences, Moscow 117218, Russia
[d] Space Research Institute, Moscow 117997, Russia

*E-mail: tnp51@yandex.ru (N.P. Topchiev)

**Abstract**

The future space-based GAMMA-400 gamma-ray telescope will operate onboard the Russian astrophysical observatory in a highly elliptic orbit during 7 years to observe Galactic plane, Galactic Center, Fermi Bubbles, Crab, Vela, Cygnus X, Geminga, Sun, and other regions and measure gamma- and cosmic-ray fluxes. Observations will be performed in the point-source mode continuously for a long time (~100 days). GAMMA-400 will measure gamma rays in the energy range from ~20 MeV to several TeV and cosmic-ray electrons + positrons up to several tens TeV. GAMMA-400 instrument will have very good angle and energy resolutions, high separation efficiency of gamma rays from cosmic-ray background, as well as electrons + positrons from protons. The main feature of GAMMA-400 is the unprecedented angular resolution for energies >30 GeV better than the space-based and ground-based gamma-ray telescopes by a factor of 5-10. GAMMA-400 observations will permit to resolve gamma rays from annihilation or decay of dark matter particles, identify many discrete sources, clarify the structure of extended sources, specify the data on cosmic-ray electron + positron spectra.

# 1 Introduction

The last few decades have been marked by a new stage in the development of astronomy, due to active observations in all ranges of the electromagnetic spectrum - from the low-frequency radio emission to the gamma-ray emission of high and ultrahigh energies. The possibility of observation in all wavelength ranges, as well as the use of neutrino and gravitational wave facilities, made it possible to implement an integrated approach in the study of astrophysical objects.

A significant contribution to astrophysical research is made by gamma-ray astronomy. Numerous studies carried out in this area have made it possible to discover galactic diffuse gamma-ray emission, isotropic extragalactic gamma-ray emission, to detect and investigate numerous galactic and extragalactic gamma-ray sources, to study gamma-ray lines, gamma-ray bursts, to study gamma-ray emission from the Sun, as well as to search for dark matter particles.

At present, the Fermi-LAT gamma-ray telescope has continued to observe the gamma-ray emission in space since 2008. The fourth Fermi-LAT catalog (Abdollahi et al., 2020) contains 5065 sources for the energy range from 50 MeV to 1000 GeV, but ~30% of gamma-ray sources are still unidentified. It should be noted that Fermi-LAT carries out observations in the scanning mode and the source observation time is only ~15% of the telescope operation time (Abdollahi et al., 2020). The ground-based facilities VERITAS, MAGIC, H.E.S.S., HAWC, ARGO-YBJ, LHAASO and others have observed the gamma-ray emission from only ~250 gamma-ray sources in the energy range above ~100 GeV (http://tevcat.uchicago.edu/). The problem of identifying gamma-ray sources with the best angular resolution in space-based experiments remains one of the central goals.

Another relevant goal is to search for and study gamma-ray bursts (GRBs). Despite the GRB searching with Fermi-LAT (Ajello et al., 2019), Fermi-GBM (von Keinlin et al., 2020), and Swift-BAT (Lien et al., 2016) the complete understanding in the physical processes responsible for high-energy emission from GRBs is not reached.

Attempts to study the properties of dark matter (DM) particles are carried out by direct and indirect detection methods. The indirect detection method consists in recording not DM particles themselves, but the products of their potential annihilation or decay: cosmic-ray (CR) charged particles, antiparticles and gamma rays (the PAMELA, AMS-2, ATIC, Fermi-LAT, DAMPE, CALET, HESS-II, MAGIC, VERITAS, HAWC, AUGER experiments), and neutrinos (the IceCube/DeepCore/PINGU, ANTARES/KM3NET, BAIKAL-GVD, Super-Kamiokande/HyperKamiokande, AUGER experiments).

Weakly interacting massive particles (WIMPs) with mass between several GeV and several TeV are still considered as the most probable candidate (e.g. Leane et al., 2018) for the role of DM.

WIMPs can annihilate or decay with the production of gamma rays and electrons + positrons. They can produce both the excess in continuous energy spectrum and narrow gamma-ray lines. Axion-like particles (ALPs) are also proposed as candidate particles for DM (e.g. Calore et al., 2020). Search for and study of the potential emission or particles from DM with the best angular and energy resolutions remains one of the central goals.

ATIC (Chang et al., 2008), Fermi-LAT (Abdollahi et al., 2017), PAMELA (Adriani et al., 2017), AMS-2 (Aguilar et al, 2014), CALET (Adriani et al., 2018), DAMPE (Ambrosi et al., 2017), MAGIC (Tridon et al., 2011), VERITAS (Staszak et al., 2015), H.E.S.S. (Aharonian et al., 2009) measured energy spectra of CR electrons + positrons in the TeV energy range. The data of experiments are not entirely consistent with each other, however, the most of observations show the existence of a remarkable flux suppression at E ~1 TeV. The existence of this cut-off has been first reported by the H.E.S.S., confirmed by MAGIC, VERITAS, and also obtained in direct measurements by DAMPE and CALET detectors. At higher energies direct measurements extended only up to ~4.8 TeV. The obtained results in the energy range above 100 GeV (e.g. Evoli et al., 2021) require further improvement up to 10 TeV.

Therefore new direct observations of gamma-ray emission in the energy range from ~20 MeV up to several TeV, as well as electron + positron fluxes up to several tens TeV are required using the space-based telescope with better angular and energy resolutions, and separation from background, in order to identify many gamma-ray sources, resolve possible gamma-ray lines or clarify an excess in continuous energy spectrum from DM, specify energy spectra of CR electrons + positrons.

## 2 The GAMMA-400 astrophysical observatory

The Russian GAMMA-400 astrophysical observatory will include the Navigator spacecraft platform (developed by Lavochkin Association, https://www.laspace.ru/projects/astrophysics/gamma-400/), the GAMMA-400 gamma-ray telescope (developed, in the main, by Lebedev Physical Institute, https://gamma400.lebedev.ru/indexeng.html, and National Research Nuclear University "MEPhI") and additional instruments (the ART-XC X-ray telescope with the energy range of 5-30 keV and magnetic plasma detectors developed by Space Research Institute of the Russian Academy of Sciences). The GAMMA-400 gamma-ray telescope is installed on a special truss at a distance of about 2 m from the Navigator space platform. The gamma- and X-ray telescopes are installed coaxially without overlapping fields of view, which are ±45° and ±0.5°, respectively (Fig. 1).

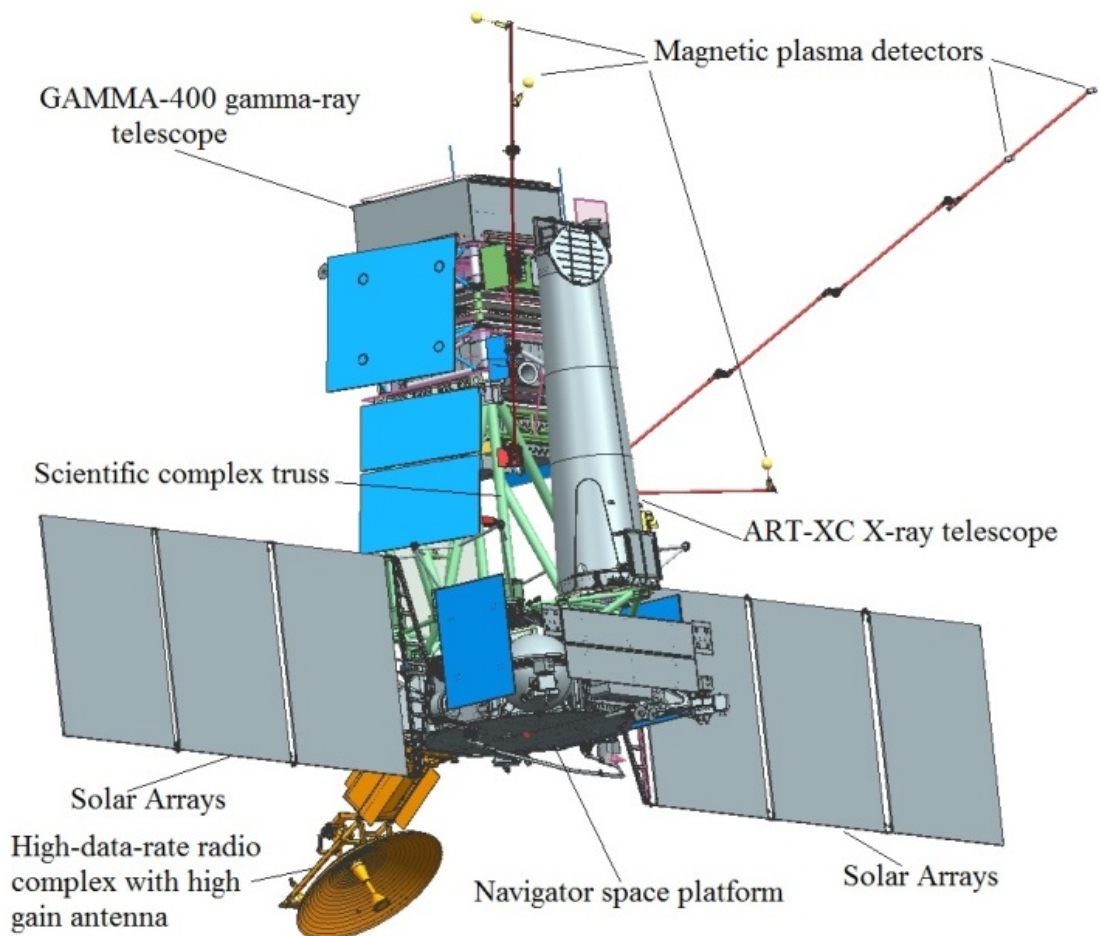

Fig. 1. The Russian GAMMA-400 astrophysical observatory, including the Navigator spacecraft platform, the GAMMA-400 gamma-ray telescope and additional instruments (the ART-XC X-ray telescope and magnetic plasma detectors).

The GAMMA-400 astrophysical observatory includes the high-data-rate radio complex (HDRRC) with a high-gain antenna (Fig. 2). HDRRC consists of antenna feeder system and on-board (airborne) radio-technical complex.

The antenna feeder system includes:
- two-mirror transmitting high-gain antenna with a diameter of 1.5 m;
- waveguide path and filters.

The on-board (airborne) radio-technical complex contains radio-transmitting device, operating at the frequency of 15 GHz.

HDRRC transmits data with the speed of 160 Mbps, using quadrature phase-shift keying modulation (QPSK) at the frequency of 15 GHz, while due to the use of error-correcting coding, an error probability is no more than $10^{-9}$.

HDRRC has two modes of operation:
- operating mode with rated output power (output signal power from 40 to 50 W);
- operating mode with reduced output power (output signal power from 4 to 7 W).

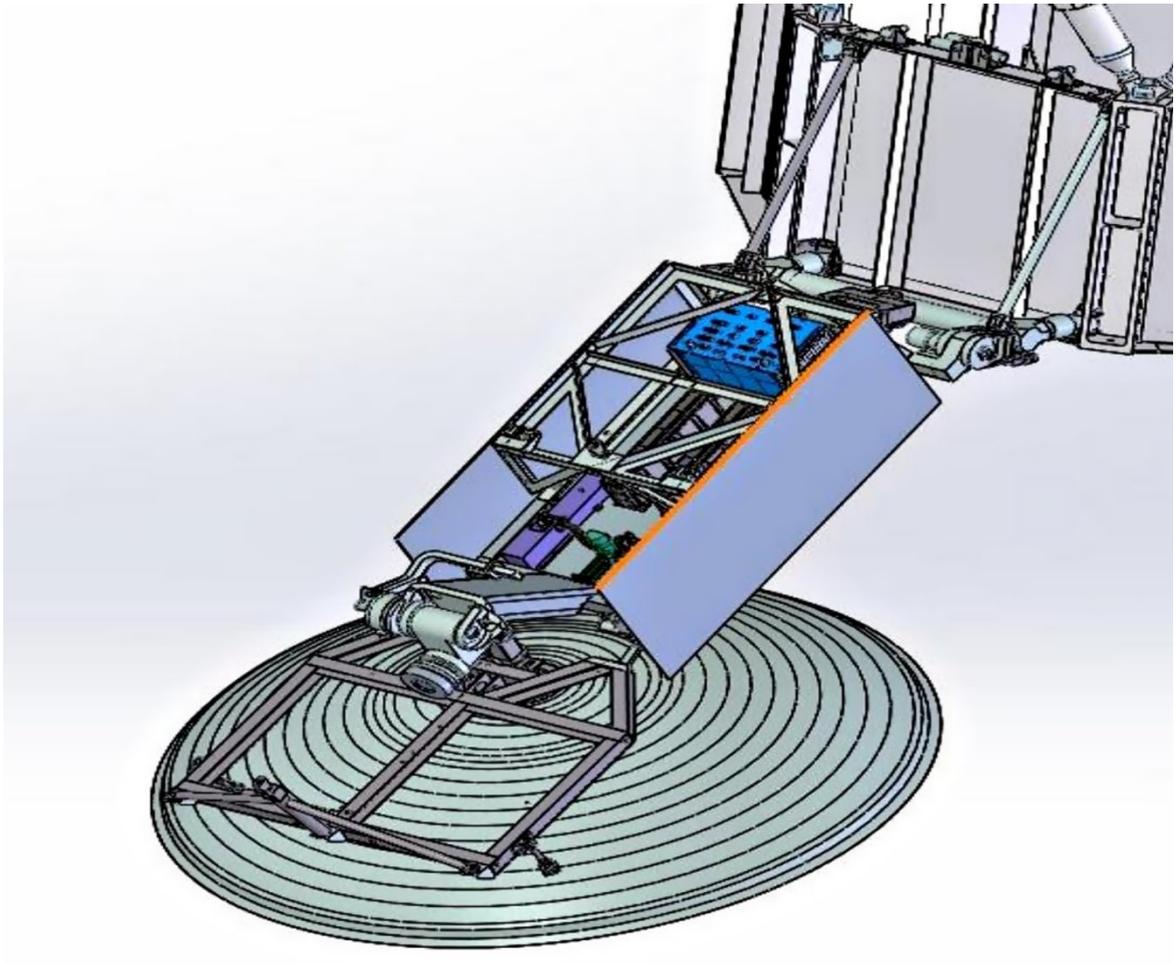

Fig. 2. High-data-rate radio complex with a high-gain antenna.

As a ground receiving station (GRS), it is proposed to use the radio-astronomy complex based on the RT-22 radio-telescope (Fig. 3) in Pushchino (Lebedev Physical Institute), the same station as for the RadioAstron mission (Spectr-R) (Kardashev et al., 2013).

GRS functions in cooperation with on-board HDRRC equipment and provides the solution of the following functions:
- tracking the spacecraft in its motion along the sky during the communication session;
- reception and decoding of the scientific data flows at the speed of 2 × 80 Mbit/s with relative error $10^{-9}$ and their recording after decoding on special digital recorders (RDR, MARK-5 and others) together with the exact local time information;
- signal selection of the carrier signal frequency of the transmission channel, and the measurement of the frequency of the residual Doppler shift and the current phase difference signal of H-maser GRS, digitization of these data with the specified rate, their recording, which is bound to the current time and the transfer to consumers.

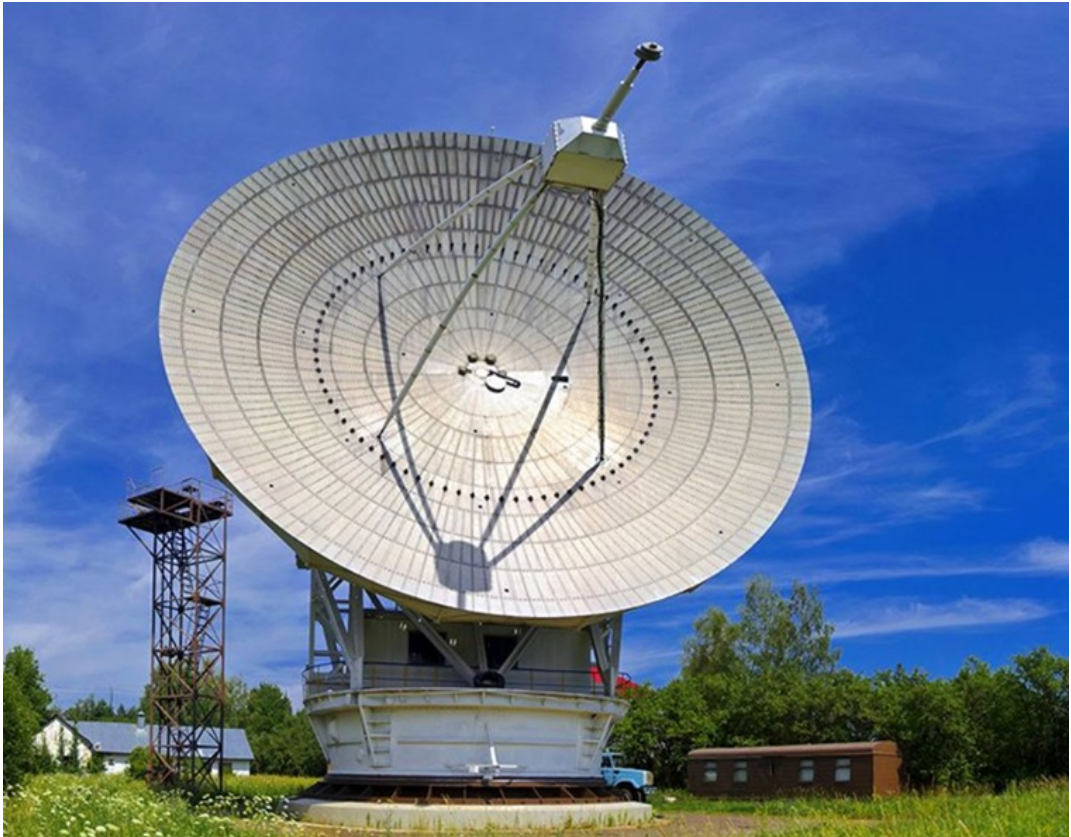

Fig. 3. The RT-22 radio-telescope in Pushchino as a ground receiving station.

Using the Navigator spacecraft platform gives the GAMMA-400 experiment a highly unique opportunity for the near future gamma- and cosmic-ray science, since it allows us to install the scientific payload (mass of ~3000 kg, power consumption of 2000 W, telemetry downlink of 100 GB/day, lifetime more than 7 years), which will provide the significant contribution of GAMMA-400 as the next generation instrument for gamma-ray astronomy and cosmic-ray physics.

The GAMMA-400 observatory will be initially launched into a highly elliptical orbit (with an apogee of 300 000 km and a perigee of 500 km, with an inclination of 51.4°) with 7 days orbital period. Under the influence of gravitational disturbances of the Sun, Moon, and the Earth after ~6 months the orbit will transform to about an approximately circular one with a radius of ~200 000 km and will not suffer from the Earth's occultation and be outside the radiation belts. A great advantage of such an orbit is the fact that the many astrophysical objects will always be available for gamma-ray astronomy, since the Earth will not cover a significant fraction of the sky, as in the case for the low-Earth orbit. Moreover, the GAMMA-400 source pointing strategy (continuous point-source observation for a long time ~100 days) will be properly defined to maximize the physics outcome of the experiment in contrast to the scanning mode for the current Fermi-LAT, CALET, DAMPE and future HERD, AMS-100 experiments. The launch of the GAMMA-400 space observatory is scheduled for 2030.

## 3 The GAMMA-400 gamma-ray telescope

### 3.1 The GAMMA-400 physical scheme

The current physical scheme of the GAMMA-400 gamma-ray telescope (Galper et al., 2013; Topchiev et al., 2017; Galper et al., 2018; Topchiev et al., 2019; Egorov et al., 2020) is shown in Fig. 4.

GAMMA-400 includes:
- anticoincidence AC top and four lateral AC lat detectors;
- converter-tracker C;

- time of flight system ToF from S1 and S2 detectors;
- two-part calorimeter from CC1 and CC2;
- four lateral detectors LD;
- preshower detector S3, shower leakage detector S4.

Electronics includes system of trigger formation (ST), scientific data acquisition system (SDAS) and other electronics units.

Figure 5 shows view of the GAMMA-400 gamma-ray telescope, where 2 star sensors (SSs) with an accuracy of better than 3" and thermal control systems (TCSs) are additionally presented.

The construction for installing systems and detectors of the GAMMA-400 gamma-ray telescope and thermal control systems are developed by Lavochkin Association. Many construction elements of the gamma-ray telescope, as well as detectors are made of carbon-filled plastic.

More detailed information about detector systems is presented in Figs. 6 and 7.

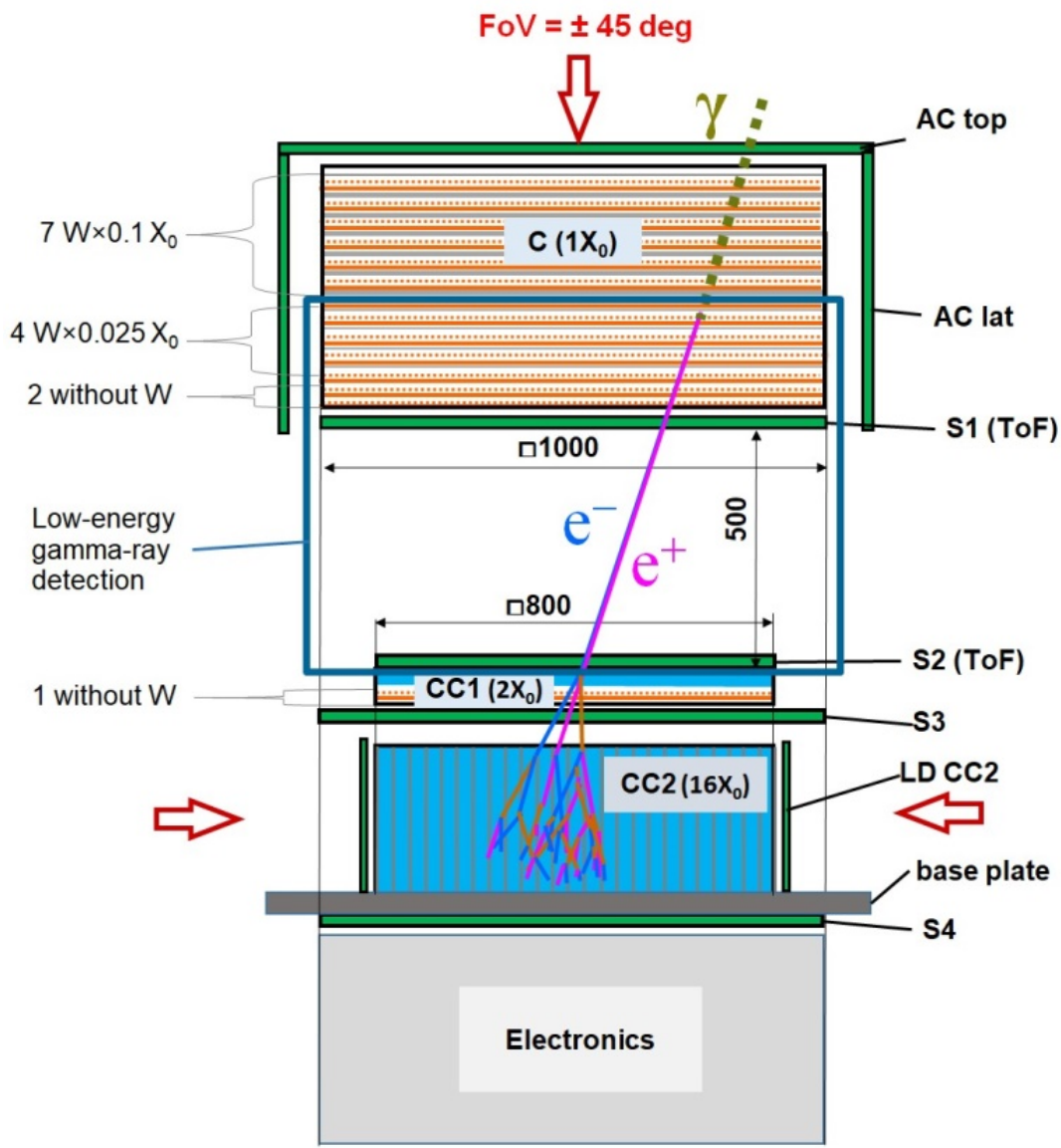

Fig. 4. The GAMMA-400 gamma-ray telescope physical scheme: anticoincidence system AC (top and four lateral detectors), converter-tracker C, time-of-flight system ToF from detectors S1 and S2, two part calorimeter from CC1 and CC2, lateral detectors LDs, preshower detector S3 and shower leakage detector S4.

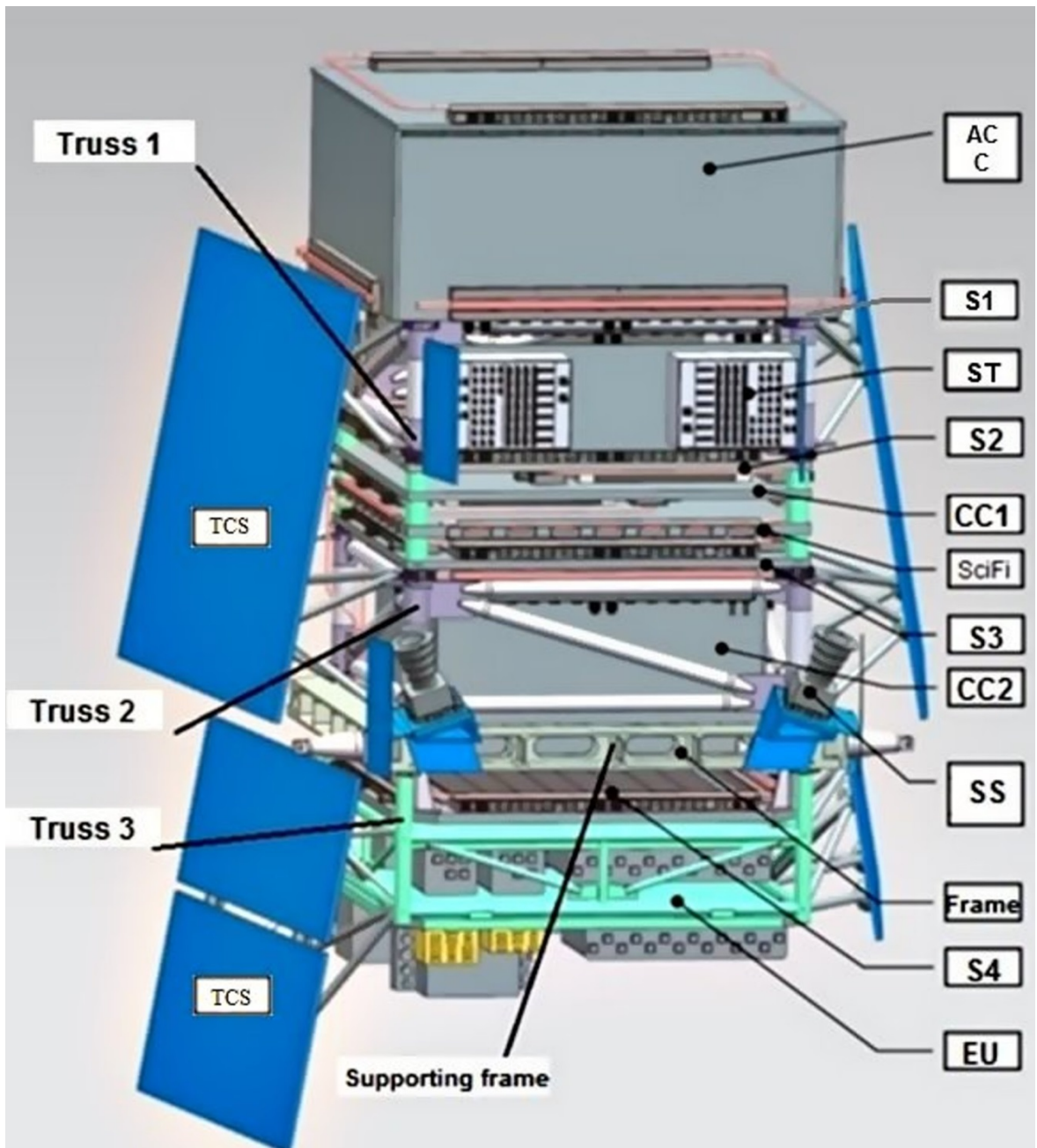

Fig. 5. View of the GAMMA-400 gamma-ray telescope.
AC is anticoincidence system, C is converter-tracker, S1 and S2 are ToF detectors, ST is system of trigger formation, CC1 and CC2 are two parts of calorimeter, SciFi is SciFi detectors in C and CC1, S3 is preshower detector, S4 is shower leakage detector, SS are star sensors, EU are electronic units, and TCS are thermal control systems.

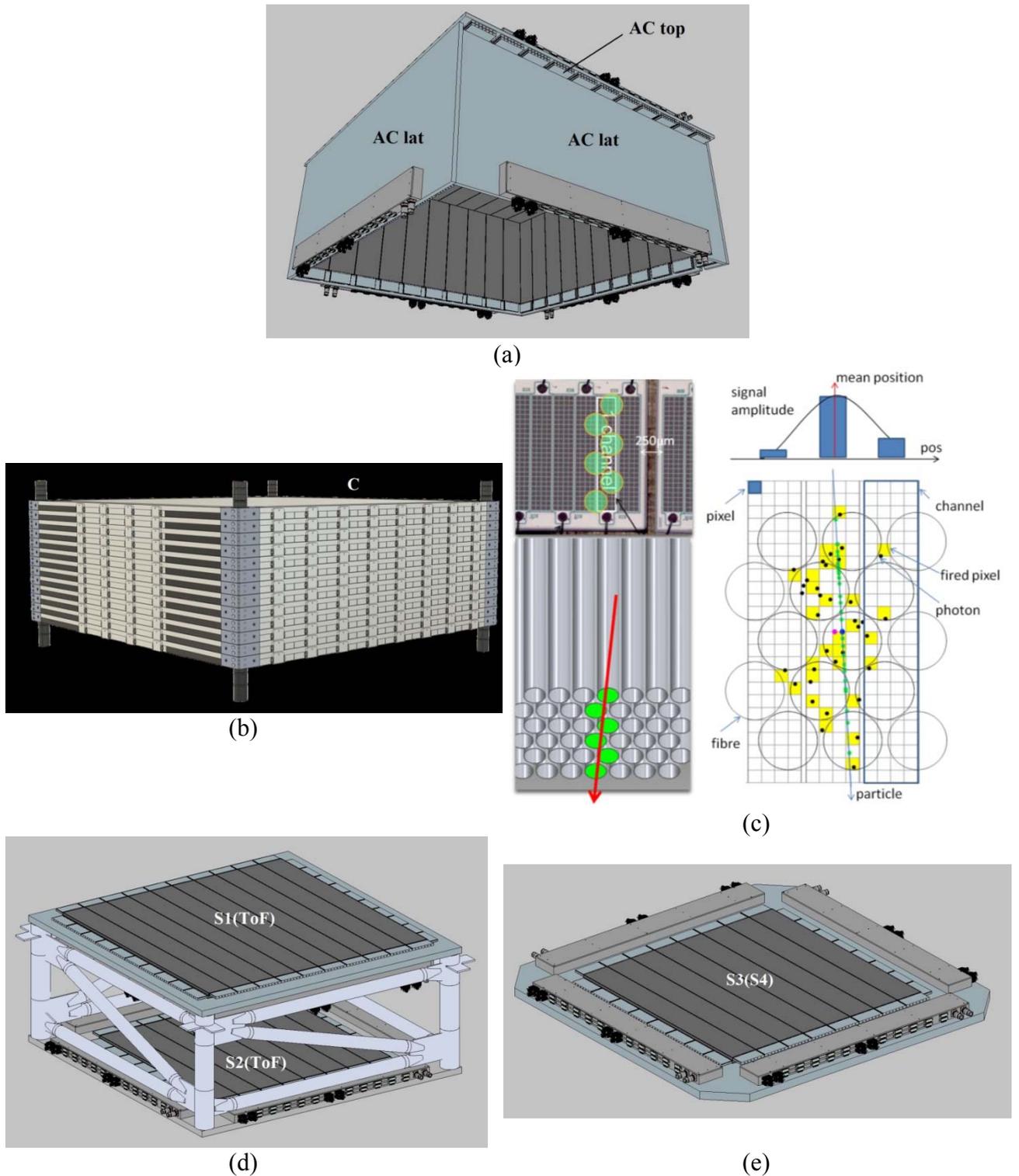

Fig. 6. The GAMMA-400 detector systems: (a) anticoincidence system AC; (b) converter-tracker C; (c) scheme of the basic principle of the SciFi tracker (Kirn et al., 2017); (d) time-of-flight system ToF from S1 and S2; (e) preshower detector S3 and shower leakage detector S4.

*Anticoincidence system*. The anticoincidence system AC is used to discriminate incoming charged particles with an efficiency of more than 0.9995 and time resolution of ~200 ps (Arkhangelskiy et al., 2020b). AC (Fig. 6a) includes one top detector AC top and four lateral detectors AC lat (located on honeycomb panels) surrounding converter-tracker C. All anticoincidence detectors are made of two layers of scintillation plastic strips with parallel orientation in both layers, which have dimensions of 10 mm in thickness for each layer and 100 mm in width. The strips of one layer are displaced by 5 mm with respect to the strips of the other layer so that there are no rectilinear slits in the system. AC top with dimensions of 1300 × 1300 mm$^2$ includes 2 layers from 22 strips with SiPM readout. AC lat with dimensions of 1300 × 500 mm$^2$ includes 2 layers from 21 strips with SiPM readout.

*Converter-tracker*. The converter-tracker is used to convert incoming gamma rays into electron-positron pair, determine conversion point and reconstruct the electromagnetic shower axis. The converter-tracker consists of 13 detector plane pairs made of scintillating fiber (SciFi) assemblies with SiPM analog readout (Fig. 6b). The SciFi assemblies contain 250 μm diameter scintillating fibers arranged in regular, hexagonal closest packed six-layer structures, with 275 μm pitch. The multichannel SiPM sensors are located at fiber ends, providing signal pulse height measurement for a subsequent light cluster barycenter calculation, resulting in singe plane coordinate resolution better than 50 μm. Scheme of the basic principle of the SciFi tracker is as follows (Fig. 6c, Kirn et al., 2017): charged particle passing through the fiber produces scintillation light, which travels to the end and is detected in a SiPM array. The black dots indicate photons arriving at the SiPM, the yellow squares indicate the pixels that fire. In each pair of planes (1050 × 1050 mm$^2$) the adjacent SciFi detectors, located on honeycomb panels, provide measurements in orthogonal directions (*x* and *y* coordinates). It is necessary to note that the technology of SciFi detectors is well developed and applied, e.g., in the LHCb tracker at Large Hadron Collider (LHC) (Kirn et al., 2017). The first 7 and next 4 pairs of planes also have tungsten (W) converter foils of 0.1 $X_0$ and 0.025 $X_0$ thick in each plane, respectively ($X_0$ is radiation interaction length) (Fig. 4). The last 2 pairs of planes have no tungsten. The part of converter-tracker with thick layers of tungsten (0.1 $X_0$) is used to detect high-energy gamma rays (above 300 MeV). The part of the converter-tracker with thin layers of tungsten (0.025 $X_0$) is used to detect low-energy gamma rays (less than 300 MeV) (Fig. 4). Total thickness of converter-tracker is ~1 $X_0$.

*Time of flight system*. Under the converter-tracker, there are two scintillation detectors S1 and S2, which form the time-of-flight (ToF) system (Fig. 6d). A distance of 500 mm between these detectors provides a sufficient flight base for effective rejection of particles coming from the lower hemisphere. Detectors S1 and S2 are installed on honeycomb panels and use SiPM readout. The S1 detector with dimensions of 1000 × 1000 mm$^2$ consists of two layers of scintillation plastic strips, each having 10 mm in thickness. There are 10 scintillation strips in each layer. The S2 detector with dimensions of 800 × 800 mm$^2$ also consists of two layers of scintillation strips, each having 10 mm in thickness. There are 8 scintillation strips in each layer. ToF is used to separate particles moving from top-down directions within the aperture of the gamma-ray telescope (determined by position of S1 and S2 detectors) of down-top directions with coefficient of separation of ~1000. The ToF time resolution is ~200 ps.

*Detectors S3 and S4*. Two-layer segmented plastic scintillation detectors S3 and S4 with dimensions of 1000 × 1000 mm$^2$ and SiPM readout are located on honeycomb panels. These detectors are necessary for improving hadronic and electromagnetic shower separation (Fig. 6e). S3 detector measures the energy release from secondary particles in shower, which is developed due to the passage of primary gamma- and cosmic-ray particles in the gamma-ray telescope and takes part in the formation of the trigger for detecting high-energy events (Leonov, et al., 2019). S4 detector measures the energy release of secondary charged particles contained in the shower leakage from the calorimeter.

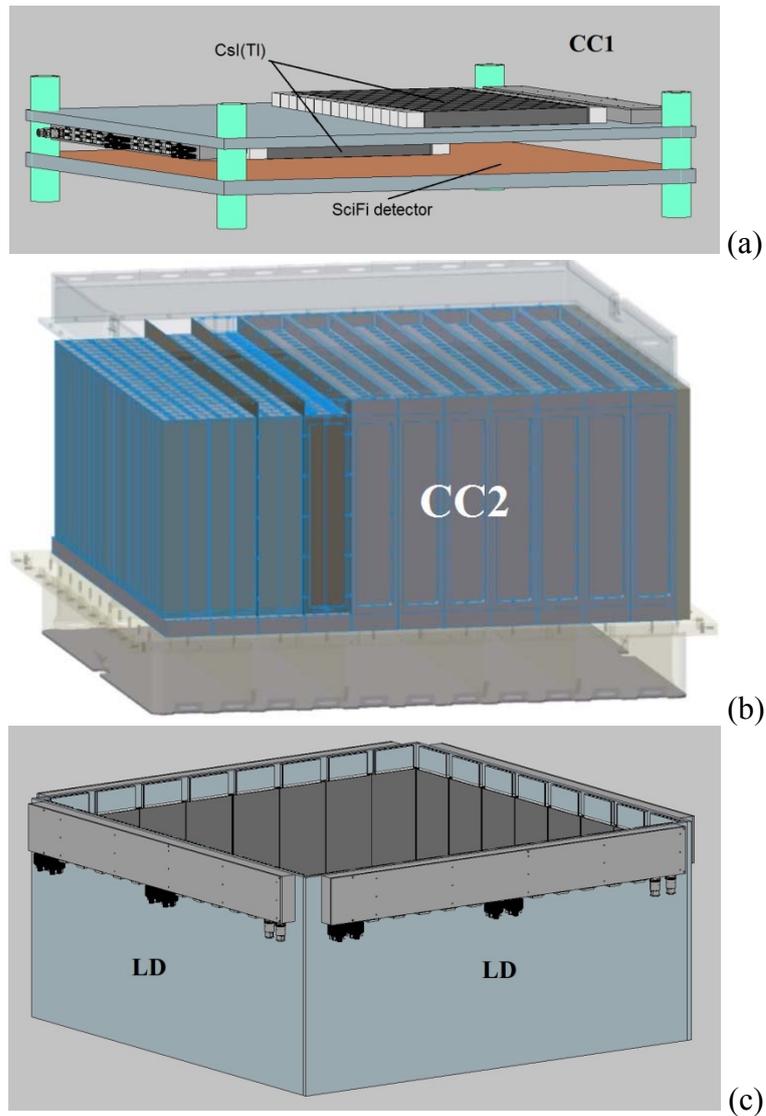

Fig. 7. Two parts of the calorimeter CC1 (a), CC2 (b) and four lateral detectors LDs (c).

*Calorimeter.* The two-part calorimeter consists of CC1 and CC2 detectors with SiPM readout. CC1 (800 × 800 mm$^2$) contains two blocks of CsI(Tl) scintillation crystals installed above and under honeycomb panel and the SciFi detector plane without W, providing *x* and *y* measurements (Fig. 7a). The SciFi detector plane is the same as last planes in the converter–tracker. CC2 (800 × 800 mm$^2$) consists of 22 × 22 matrix from CsI(Tl) crystals (Fig. 7b). Each crystal has transverse dimensions of 36 × 36 mm$^2$. CsI(Tl) crystal columns of CC2 are installed inside the cells of the carbon fiber honeycomb structure, having 0.4 mm in thickness. Total thickness of the CC1 and CC2 calorimeter is ~18 $X_0$ (~0.9 $\lambda_0$) and ~43 $X_0$ (~2.0 $\lambda_0$) for vertical and lateral particle detection, respectively ($\lambda_0$ is hadronic interaction length).

*Lateral detectors.* Four lateral scintillation plastic detectors LDs (800 × 400 mm$^2$) with SiPM readout located on honeycomb panels around of the CC2 calorimeter are used for detecting particles from lateral directions (Fig. 7c).

*System of trigger formation.* The system of trigger formation ST generates triggers for detecting various types of events analyzing the output signals of the front-end electronics of the gamma-ray telescope detectors. To improve reliability, the system is designed using a dual redundancy crossover scheme; data exchange highways are also duplicated, with each highway having its own transceiver nodes.

*Scientific data acquisition system.* The scientific data acquisition system SDAS is intended for collecting scientific data from detectors, temporary storage and transmission of scientific and service information of the gamma-ray telescope to the high-data-rate radio complex for subsequent transmission to a ground receiving station, as well as for direct control of the gamma-ray telescope. SDAS is a monoblock consisting of redundant functional modules. The choice of the configuration of modules is carried out by means of commands coming from the on-board control complex of the Navigator platform.

*Principle of operation.* Primary gamma rays pass without interaction the anticoincidence system and are converted into electron-positron pair in the converter-tracker. Further, electron-positron pair passes through the rest part of converter-tracker, time-of-flight system, preshower detector S3 and induces an electromagnetic shower in CC1 and CC2 calorimeter following by shower leakage detector S4. The energy deposits in converter-tracker, S1, S2, CC1, CC2, S3, S4, LD detectors are measured.

The responses of the gamma-ray telescope detectors AC, ToF and S3 contribute to the fast on-board trigger logic in the top-down telescope aperture. The readout electronics for fast on-board trigger and time of flight system of the GAMMA-400 gamma-ray telescope constitutes the multiprocessor structure, which collects data from the gamma-ray telescope detectors and produces the trigger decision for each event. The use of the flexible distributed system provides possibility for adaptive and operational control of the parameters of the gamma-ray telescope recording modes, and the optimization of a search for a specific physics channel according to the observational mode. In order to increase the reliability, the system is crossover and double redundant. Besides, the components of selected electronics components are space qualified and rad-hard or rad-tolerant.

The incidence direction of particles is determined by best determination of conversion point and reconstruction of the electromagnetic shower axis, using SciFi coordinate detectors in the converter-tracker and CC1.

After interaction of incident gamma rays with the GAMMA-400 detector matter backscattering secondary particles (mainly, 1-MeV photons) are arisen. These secondary particles can interact in AC and produce veto signal, excluding primary gamma rays (Fig. 8). To suppress this effect all scintillation detectors are constructed as two-layer segmented layers of strips and additional trigger logic for detecting high-energy particles is organized (Leonov, et al., 2019). Besides, the timing analysis of signals produced in AC, S1 (ToF), S2 (ToF), S3 detectors are used (Arkhangelskiy et al., 2020a).

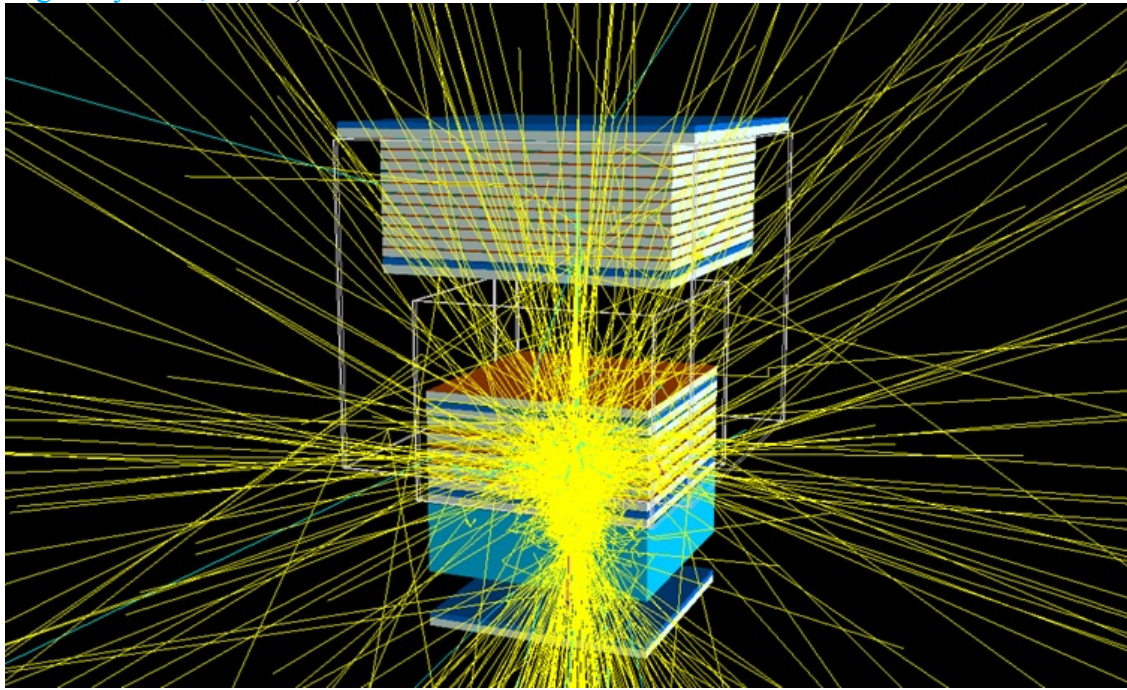

Fig. 8. Backscattering secondary particles (mainly, 1-MeV photons), which can produce veto in AC and exclude primary gamma rays.

The GAMMA-400 energy range for gamma-ray studies is from ~20 MeV to several TeV and for CR electrons + positrons is from ~1 GeV to several tens TeV. The energy of low-energy gamma rays is measured in the converter-tracker, S1 (ToF), S2 (ToF) after the conversion of gamma rays in the last six planes (Fig. 4, four planes with 0.025 $X_0$-tungsten and two without tungsten). The maximum field of view (FoV) for detecting particles from top-down directions is ±45°. The main trigger of GAMMA-400 is the following:

$$(\overline{AC} \times ToF) | (S3_{THRESHOLD} \times ToF) \qquad (1),$$

where $ToF = S1 \times S2 \times [\text{time}_{S1} < \text{time}_{S2}]$, $S3_{THRESHOLD}$ is a special energy threshold (Leonov, et al., 2019).

The value of $S3_{THRESHOLD}$ is determined from the following considerations. For triggered events with gamma rays of energy more than ~10 GeV, backscattering secondary particles (mainly, 1-MeV photons) had begun to create a veto signal in AC. To eliminate these events we construct the distributions of energy deposits in S3 detector for 10-GeV gamma rays from vertical direction, which generate veto signal in AC, and for proton spectrum with power index -2.7 with energies more than 10 GeV and introduce $S3_{THRESHOLD}$ (Fig. 9). The location of the threshold for signal in S3 detector is determined in view of trade-in between proton rejection from gamma channel and recording efficiency for high-energy gamma rays associated with backsplash. When ground processing we will additionally eliminate events with backscattering secondary particles due to reconstructing track and the time consequence of signals produced in AC, S1 (ToF), S2 (ToF).

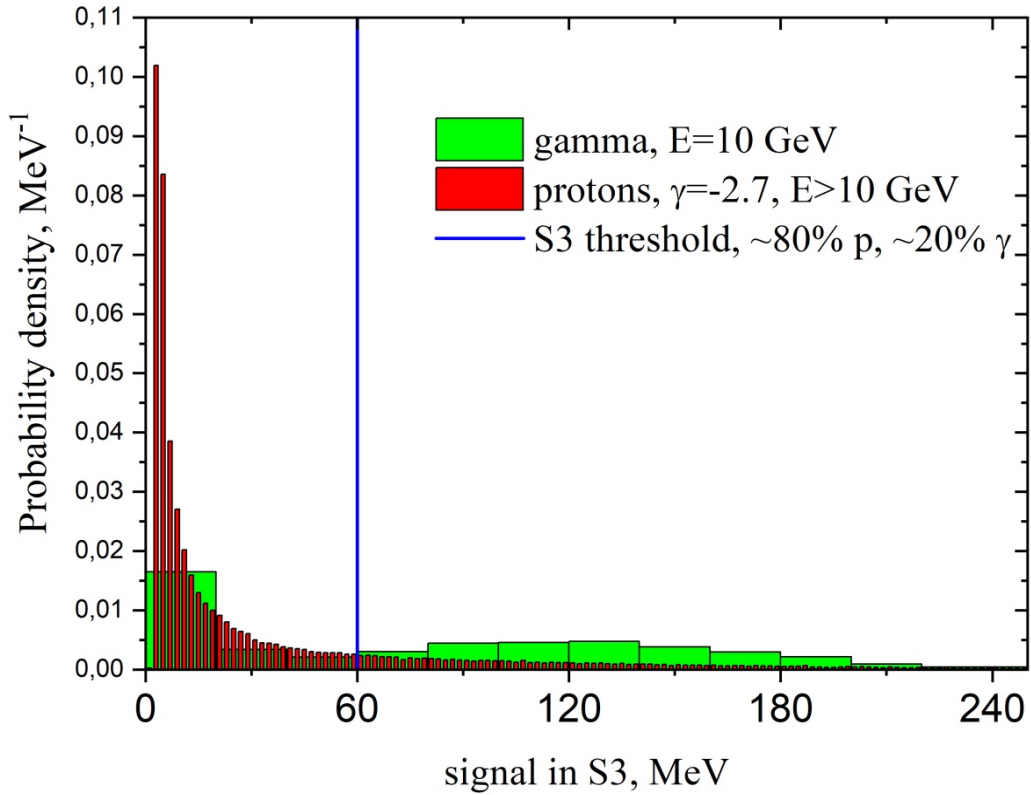

Fig. 9. The distributions of energy deposits in S3 detector for 10-GeV gamma rays from vertical direction, which generate veto signal in AC, and for proton spectrum with power index -2.7 and with energies more than 10 GeV (Leonov, et al., 2019).

In Fig. 10, the dependences of recording efficiency for gamma rays and protons from the value of signal in S3 detector are presented. The optimal value for the cutoff of the signal in S3 detector is ~60 MeV. For this cutoff, ~20% of gamma rays, which generate veto signal in AC, and ~80% of protons are rejected. In Figs. 9 and 10, the location of this energy cut is shown by blue line.

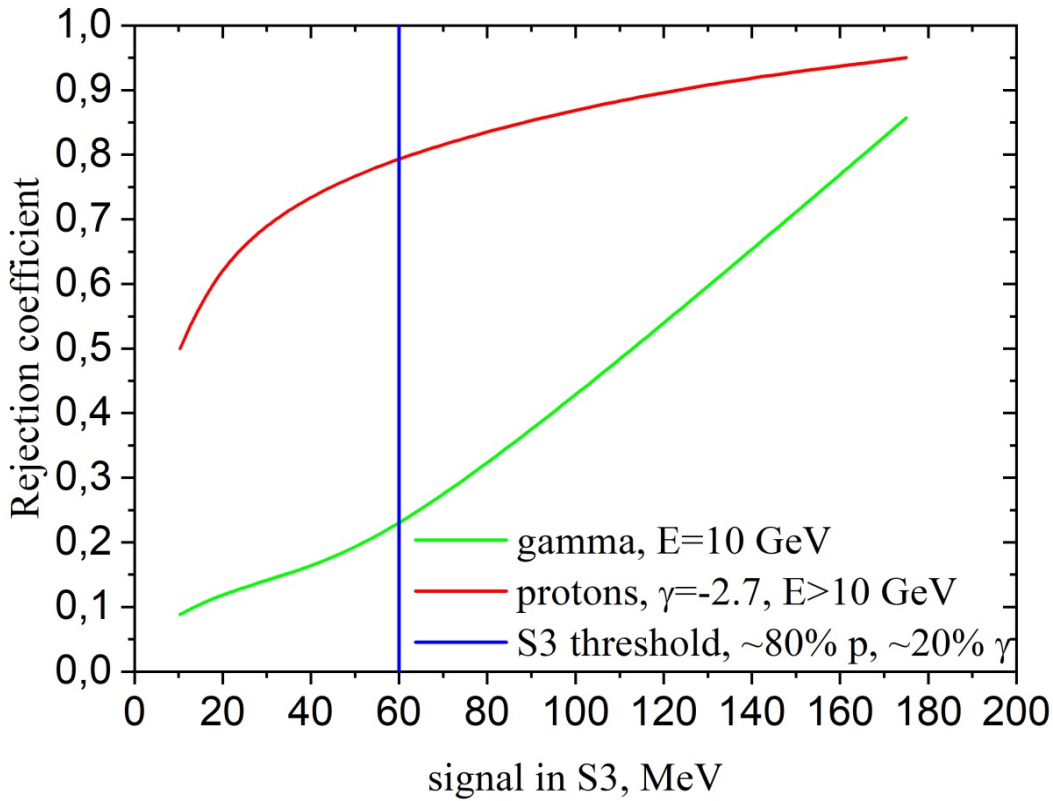

Fig. 10. The dependences of recording efficiency for gamma rays and protons from the value of signal in S3 detector (Leonov, et al., 2019).

The trigger (1) is used for detecting low- and high-energy gamma rays. For detecting CR electrons + positrons the trigger ($S3_{THRESHOLD} \times ToF$) is used. Moreover, GAMMA-400 can detect gamma and cosmic rays from lateral directions. For this, the trigger $\overline{LD} \times \overline{S_3} \times \overline{S_4} \times CC_2$ and $LD \times CC_2$ are used, respectively.

The functional diagram of the GAMMA-400 gamma-ray telescope is presented in Fig. 11 (Arkhangelskiy et al., 2020a). The module structure of the GAMMA-400 gamma-ray telescope includes following subsystems and detecting units:

- scientific measuring subsystems (SMS): converter-tracker C; anticoincidence system AC; calorimeter CC consisting of CC1 and CC2; detectors S3, LD and S4; time-of-flight system ToF consisting of S1 and S2; system of triggers formation ST;

- service subsystem, including two-star sensors and scientific complex telemetry system for cyclic registration up to 65535 housekeeping parameters of scientific complex;

- scientific data acquisition system SDAS for acquisition and pre-processing data from SMS and service subsystem; storage the data in non-volatile mass memory (1 TByte), scientific data and telemetry transfer into high-speed scientific radio line (up to 400 Mbit/sec) for their transmission to the ground receiving station and control information reception from spacecraft on-board control system via multiplexed channel MIL-STD-1553B; data decoding and transfer into SMS;

- one-time commands and power supply system OCPSS, providing power supply for scientific complex, one-time pulse radio commands sharing and their transmitting to SMS and transit of the most important telemetry parameters directly to the satellite on-board telemetry system.

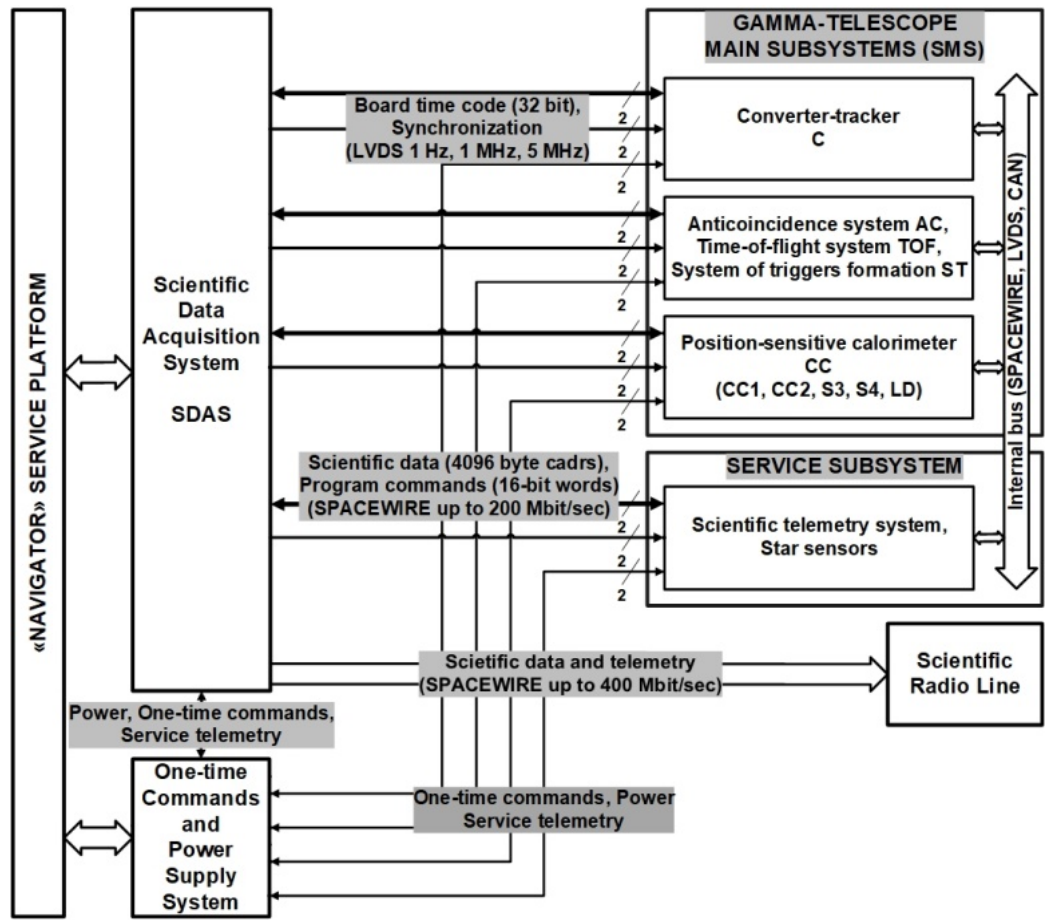

Fig. 11. The functional diagram of the GAMMA-400 gamma-ray telescope.

### 3.2 The GAMMA-400 performance

Simulations of the GAMMA-400 gamma-ray telescope performance were carried out using the GEANT4.10.01.p02 software package. An example of simulation of 50-GeV gamma-ray energy release in the gamma-ray telescope detectors is shown in Fig. 12, where detectors of the gamma-ray telescope are shown as colored, if the response exists. The incidence of gamma ray is coaxial with the telescope axis. In this figure, two projections of the gamma-ray telescope response are shown at the left and right of the figure. At the center, a view from above CC2 is shown together with anticoincidence system (AC lat) detectors.

As a result of calculations, at which it is necessary to analyze the time consequence of signals produced in S1 (ToF), S2 (ToF) taking into account time resolution obtained from test beam calibration (Arkhangelskiy et al., 2020b) and to be able reconstructing the particle track and restoring the particle energy, we obtained the following dependences:

1) The on-axis effective area vs the energy (Fig. 13). It is seen from Fig. 13 that when using only trigger ($\overline{AC} \times ToF$) the effective area begins to decrease at the energy more than 10 GeV due to backscattering particles. Using the trigger ($\overline{AC} \times ToF)|(S3 \times ToF)$ provides the effective area of ~4000 cm$^2$ up to several TeV. If the additional analysis is introduced, providing possibility to reconstruct track and to restore energy, the estimation of effective area ($S_{eff}$) decreases up to ~3200 cm$^2$.

2) The effective area vs the incident angle for the energies of 1, 10, 100 GeV using the trigger $(\overline{AC} \times ToF)|(S3 \times ToF)$ and providing possibility to reconstruct track and to restore energy (Fig. 14).

3) The differential point-source sensitivity for 1.5 and 10 year observations of the Galactic center (l = 0°, b = 0°) was estimated (Fig. 15). For this purpose we used the typical detection criterion:

$$TS = 2ln(L_1/L_0) = 25, \qquad (2)$$

where $L_1$ is the likelihood function assuming the source presence on top of background inside PSF and $L_0$ contains only the background. Similarly to the approach used by Fermi-LAT collaboration we also implied the additional detection criterion that each energy bin must have at least 10 photons from the source. Substituting the Poisson likelihood in (2) we would have:

$$TS = 2(nln(\lambda_1/\lambda_0) - \lambda_1 + \lambda_0), \qquad (3)$$

where $\lambda_1$ is the mean expected number of photons from source + background, $\lambda_0$ is that from the background only and $n$ is the measured number of photons assumed to be $n \equiv \lambda_1$. The expected photon numbers were calculated through the following way:

$$\lambda_{0i} = \int_{E_{il}}^{E_{iu}} \int_{E\prime-3\sigma_E(E\prime)}^{E\prime+3\sigma_E(E\prime)} \int_{\Omega_{PSF}(E)} \varepsilon(E) I_b(E) \frac{1}{\sqrt{2\pi}\sigma_E(E)} exp\left(-\frac{(E\prime-E)^2}{2\sigma_E(E)^2}\right) cos(\theta) d\Omega dE dE', \qquad (4)$$

$$\lambda_{1i} = \int_{E_{il}}^{E_{iu}} \int_{E\prime-3\sigma_E(E\prime)}^{E\prime+3\sigma_E(E\prime)} \varepsilon(E) \left(k\frac{c_i}{E^2} + \int_{\Omega_{PSF}(E)} I_b(E) cos(\theta) d\Omega\right) \frac{1}{\sqrt{2\pi}\sigma_E(E)} exp\left(-\frac{(E\prime-E)^2}{2\sigma_E(E)^2}\right) dE dE'. \qquad (5)$$

In the equations above, $E_{il}$ and $E_{iu}$ denotes lower and upper margins of the $i$th energy bin (we choose 4 energy bins per decade, see Fig. 13); $\varepsilon(E) = \int A(E, \theta(t)) dt$ is the exposure of the relevant point in the sky; $I_b(E)$ is the background intensity in that point, which include both the diffuse Galactic component and isotropic background. The spectra of these backgrounds were taken from Fermi-LAT data at https://fermi.gsfc.nasa.gov/ssc/data/access/lat/BackgroundModels.html using Aladin software (https://aladin.u-strasbg.fr/aladin.gml). The Galactic background was averaged in the square 0.5°x0.5° around the relevant point and then assumed isotropic inside our PSF. Indeed this is a simplified methodology, which may introduce some uncertainties. $\Omega_{PSF}(E)$ is the solid angle size of PSF. We choose this size to have 68% containment of the source photons. This is taken into account by the coefficient $k = 0.68$ in (5). The Gaussian function in (4)-(5) reflects the effect of finite energy resolution of the telescope, i.e., the energy dispersion. However, overall dependence of the sensitivity on $k$ and energy resolution is weak. We neglected by the effect of source confusion and assumed a typical inverse square spectral shape for point sources. Then we substituted (4)-(5) into (2)-(3) and solved for $c_i$. We defined the minimal detectable differential flux $m_i$ as follows:

$$m_i = \int_{E_{il}}^{E_{iu}} \frac{c_i}{E^2} E dE. \qquad (6)$$

This quantity physically means the energy flux from the source inside $i$th energy bins. The result of calculation is presented in Fig. 15 for the Galactic Center and two exposure times of 1.5 and 10 years (continuous exposure in the center of FoV). The curves reach their minima, where the transition happens from TS = 25 to >10 photons/bin conditions as driving criteria. In other words, above ~10 GeV the backgrounds gradually cease, and the scarcity of photons from source becomes the key factor, which limits sensitivity.

4) The angular resolution vs the energy in comparison with Fermi-LAT, HERD, DAMPE, and CTA (https://www.cta-observatory.org/science/ctao-performance/#1472563318157-d0191bc5-0280) (Fig. 16). The angular resolution for $E_\gamma$ = 100 GeV is ~0.01°.

5) The on-axis energy resolution vs the energy in comparison with Fermi-LAT, HERD, DAMPE, and CTA (https://www.cta-observatory.org/science/ctao-performance/#1472563318157-d0191bc5-0280) (Fig. 17). The energy resolution for $E_\gamma$ = 100 GeV is ~2%.

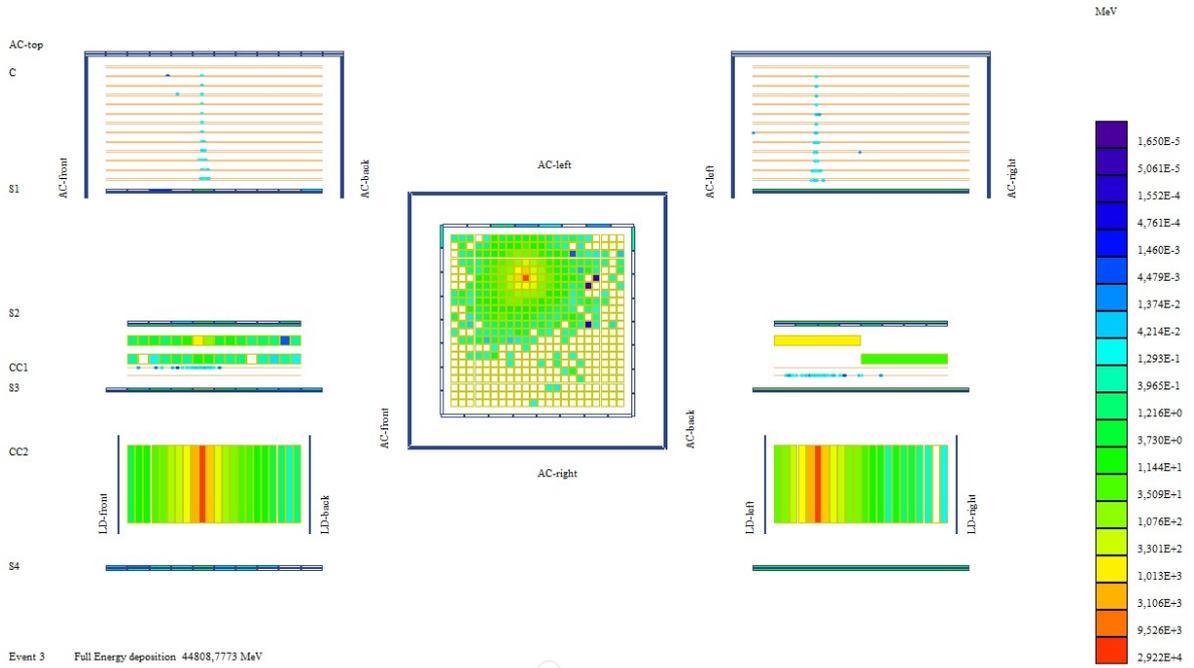

Fig. 12. An example of simulation of energy release in the gamma-ray telescope detectors for 50-GeV gamma ray.

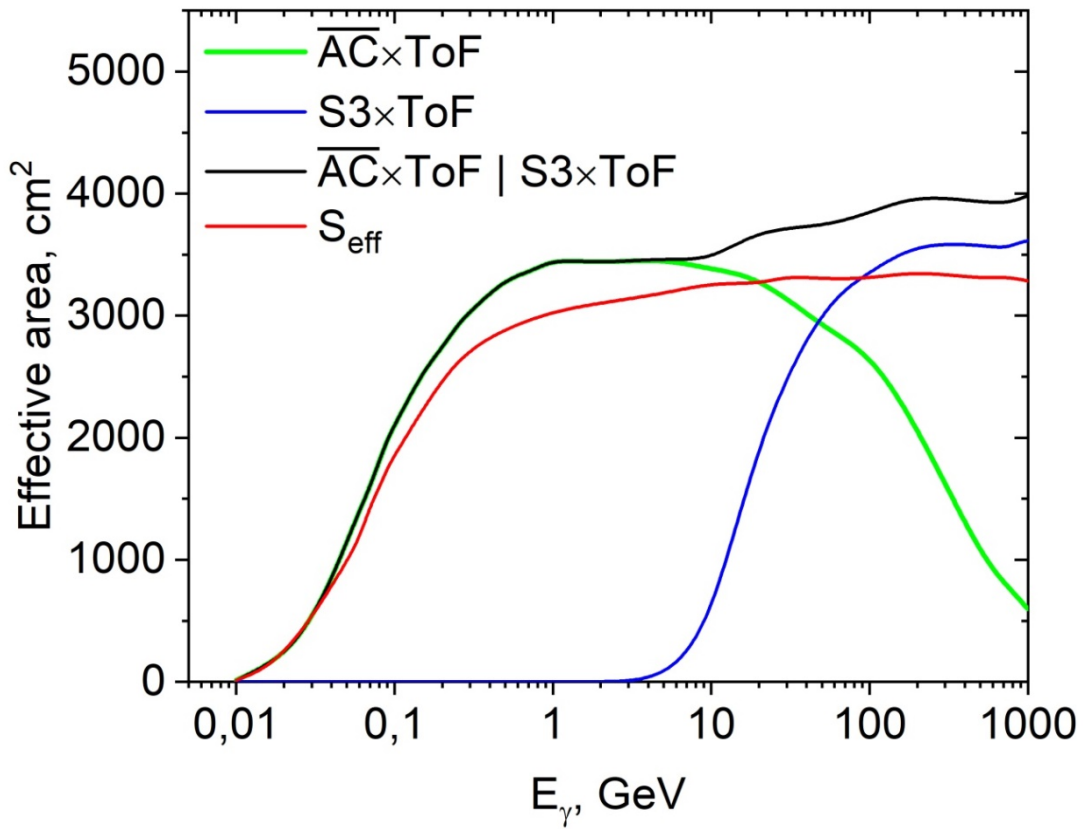

Fig. 13. Dependence of the GAMMA-400 on-axis effective area vs the energy using triggers $\overline{AC} \times ToF$ (green), $S3 \times ToF$ (blue), $\overline{(AC \times ToF)}|(S3 \times ToF)$ (black), as well as trigger $\overline{(AC \times ToF)}|(S3 \times ToF)$ and providing possibility to reconstruct track and to restore energy (red).

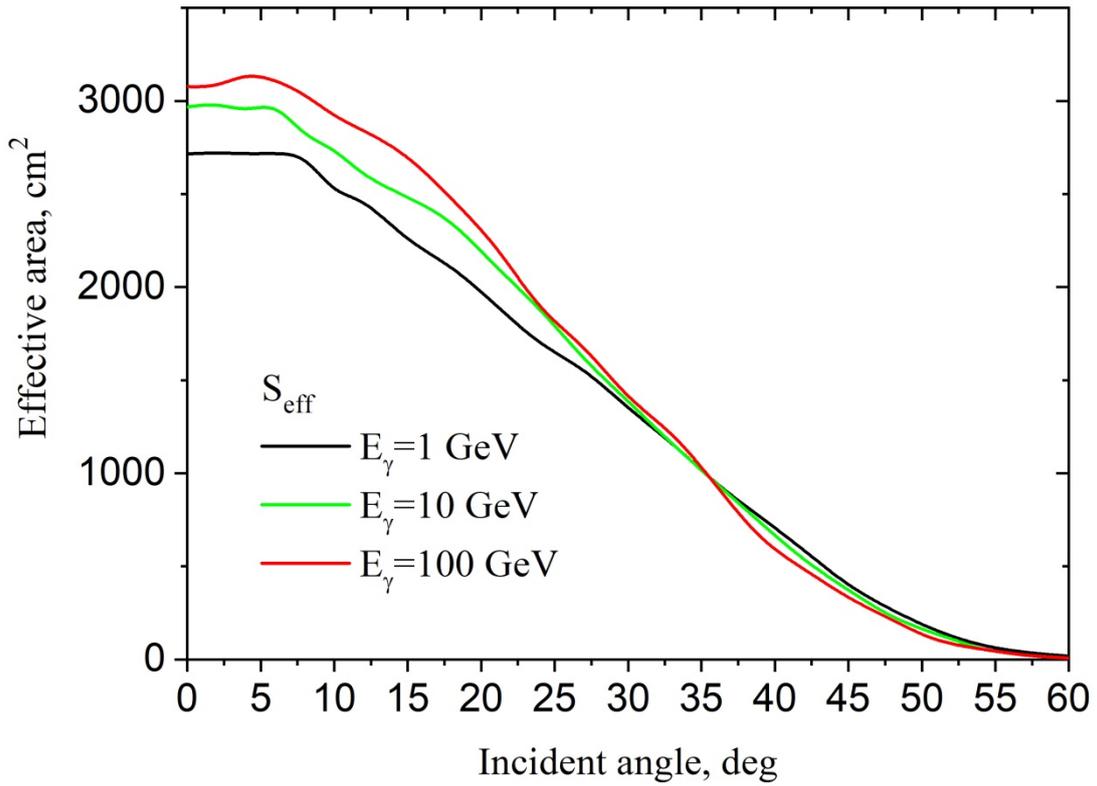

Fig. 14. Dependence of the GAMMA-400 effective area vs the incident angle for the energies of 1 (black), 10 (green), 100 (red) GeV using trigger $(\overline{AC} \times ToF)|(S3 \times ToF)$ and providing possibility to reconstruct track and to restore energy.

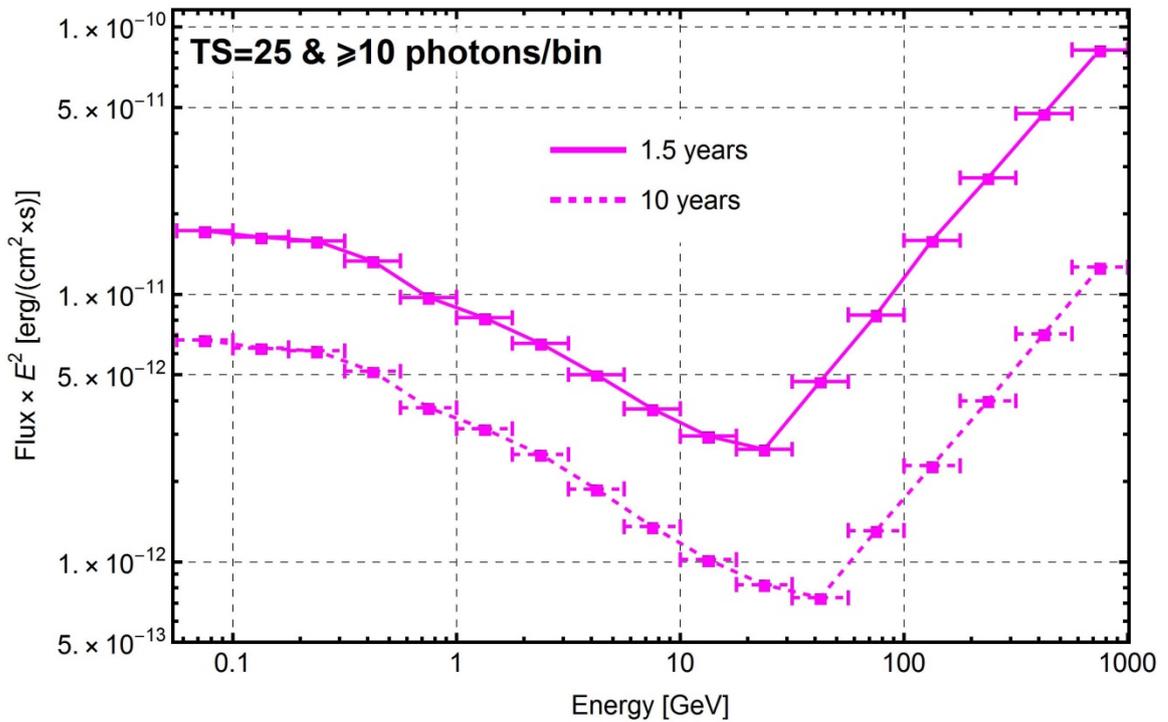

Fig. 15. The GAMMA-400 differential point-source sensitivity for 1.5 and 10 year observations of the Galactic center (l = 0°, b = 0°). The horizontal lines reflect the chosen energy bins for flux density integration.

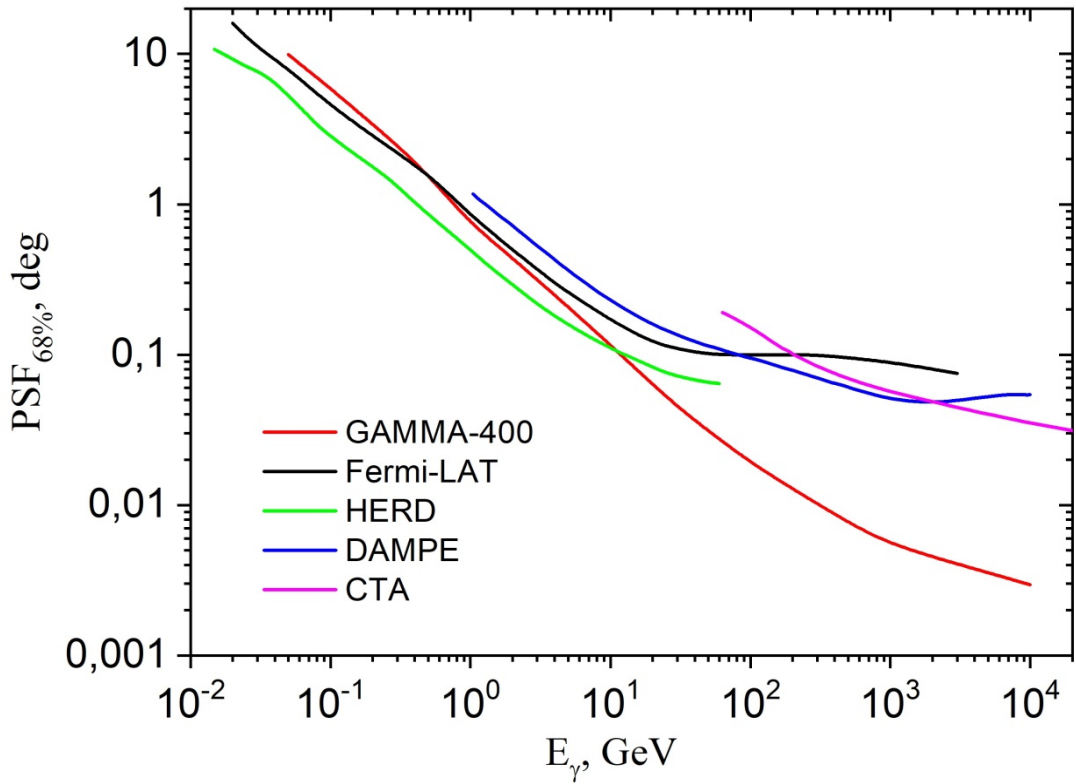

Fig. 16. Dependence of the GAMMA-400 angular resolution vs the energy in comparison with Fermi-LAT, HERD, DAMPE, and CTA.

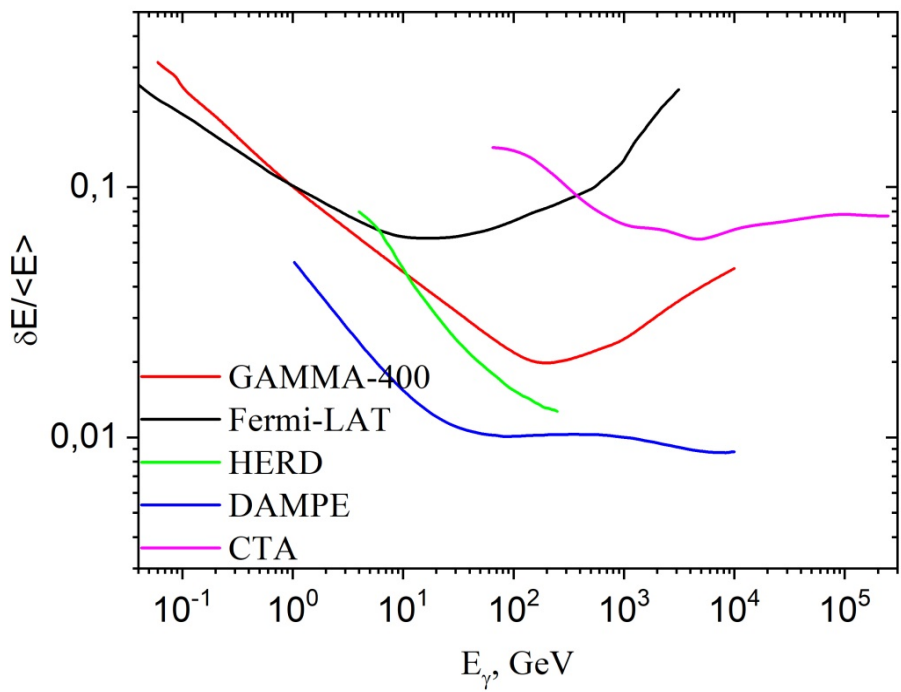

Fig. 17. Dependence of the GAMMA-400 on-axis energy resolution vs the energy in comparison with Fermi-LAT, HERD, DAMPE, and CTA.

Separation of gamma rays against background is carried out using high-efficiency AC system (more than 0.9995, Arkhangelskiy et al., 2020b), high coefficient of separation of events coming from top-down and down-top directions of ~1000 in the time-of-flight system, as well as separation of electromagnetic and hadronic showers in calorimeters and S4 (~$10^4$) (Leonov et al., 2015). For gamma rays the rejection is ~$10^5$ due to the additional analysis of the fired strips of AC detectors located along the restored trajectory of initial particle.

GAMMA-400 will detect CR electron + positron fluxes from top-down directions and from lateral directions around the calorimeter. Examples of energy distribution for 100-GeV electron incident on the center of the gamma-ray telescope along the vertical axis (Fig. 18) and on the center of one of the lateral detectors (Fig. 19). For these cases, the energy resolutions are better than 2% and 1%, respectively. Table 1 shows a comparison of performance for GAMMA-400 (top-down and lateral aperture), Fermi-LAT (Atwood et al., 2009), PAMELA (Adriani et al., 2014), AMS-2 (Aguilar et al., 2013), CALET (Mori et al., 2013), DAMPE (Chang et al., 2017), HERD (Cattaneo et al., 2019). It is seen that due to large calorimeter area (0.7 m$^2$ for top-down and 1.0 m$^2$ for lateral aperture) and calorimeter thickness (18 $X_0$ for top-down and 43 $X_0$ for lateral aperture) GAMMA-400 can improve electron + positron data statistics.

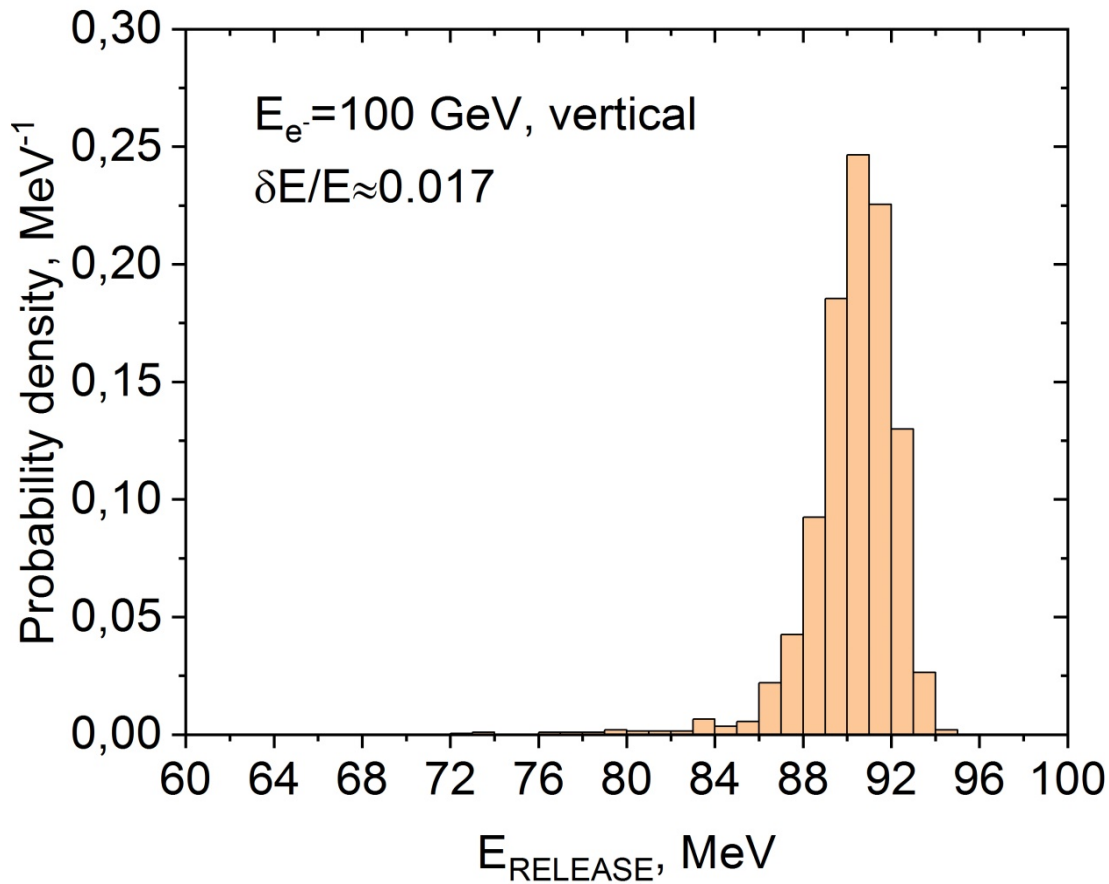

Fig. 18. An example of energy distribution for 100-GeV electron incident on the center of the gamma-ray telescope along the vertical axis. In this case, the energy resolution is better than 2%.

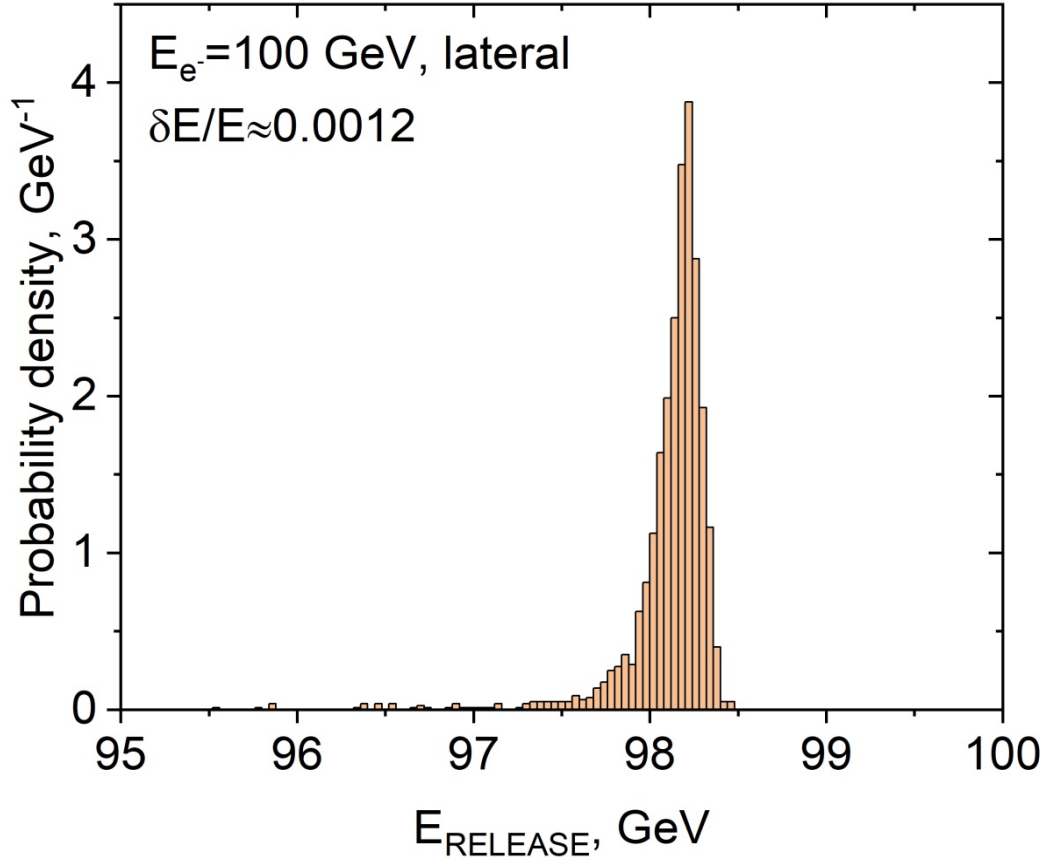

Fig. 19. An example of energy distribution for 100-GeV electron incident on the center of one of the lateral detectors. In this case, the energy resolution is better than 1%.

Table 1. Comparison of performance when detecting electrons + positrons for GAMMA-400 (top-down and lateral aperture), Fermi-LAT, PAMELA, AMS-2, CALET, DAMPE, HERD

|  | GAMMA-400 | | Fermi-LAT | PAMELA | AMS-2 | CALET | DAMPE | HERD |
|---|---|---|---|---|---|---|---|---|
| Aperture | top-down | 4 sides | top-down | top-down | top-down | top-down | top-down | 5 sides |
| Acceptance, $m^2$ sr | ~0.3 ($E_e$ = 100 GeV) | ~0.5 ($E_e$ = 100 GeV) | 2.5 | 0.02 | 0.4 | 0.1 | 0.3 | 3 |
| Proton rejection factor | ~$10^4$ | ~$5 \times 10^3$ | ~$10^4$ | ~$10^4$ | ~$10^4$ | $10^5$ | $10^5$ | >$10^5$ |
| Calorimeter area, $m^2$ | 0.7 | $4 \times 0.24$ | 0.85 | 0.06 | 0.42 | 0.1 | 0.36 | $5 \times 0.4$ |
| Calorimeter thickness, $X_0$ | 18 | 43 | 8.6 | 16 | 16 | 30 | 32 | 55 |

Figure 20 shows the GAMMA-400 acceptance vs energy when detecting electrons + positrons for top-down directions with the trigger ($S3 \times ToF$) (black) and in addition with separation criterion of electrons + positrons from protons, which will be performed when ground processing (red). The separation criterion of electrons + positrons from protons allows us to identify events with energy leakage from the calorimeter, especially using response of S4 detector, as described further.

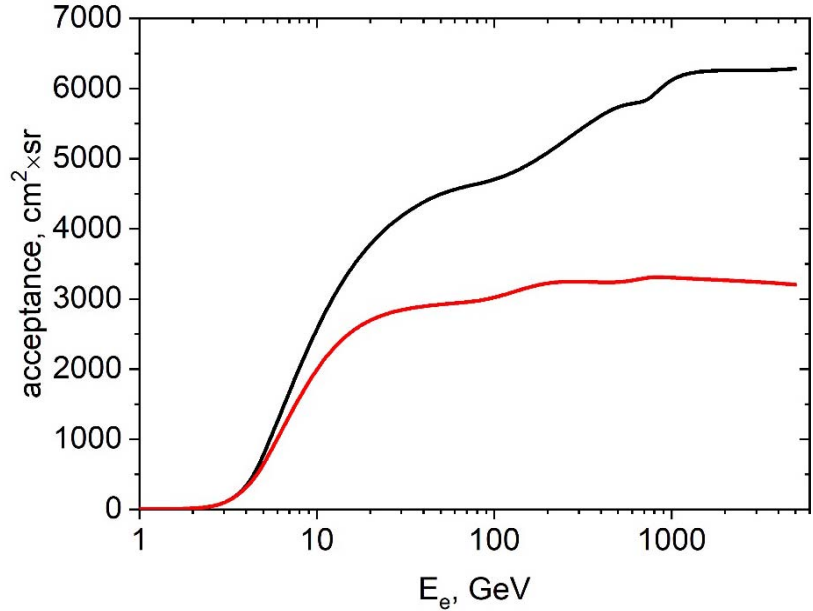

Fig. 20. Dependence of the GAMMA-400 acceptance vs energy when detecting electrons + positrons for top-down directions with the trigger ($S3 \times ToF$) (black) and in addition with separation criteria of electrons + positrons from protons, which will be performed when ground processing (red).

The additional separation provides the possibility to restore the energy and consists of the following. Using the combined information from all GAMMA-400 detector systems, it is possible to reject electrons from protons. The methods to separate electrons from protons presented in (Leonov et al., 2015; Leonov et al., 2019) are based on the difference of the development of hadronic and electromagnetic showers inside the gamma-ray telescope.

To take into account the fact that the hadronic cascade begins to develop deeper inside the gamma-ray telescope than the electromagnetic one, the signals in S1 and S2 are considered. The difference in electron and proton distributions provides the rejection factor ~3.

Additional rejection is obtained when analyzing the CC2 signals. The first criterion is based on the difference of the transversal size for hadronic and electromagnetic showers. Analyzing the distributions of protons and gamma (electrons) for RMS in CC2, it is possible to obtain a rejection factor ~15. The second criterion concerns the distribution of energy release in the hadronic and electromagnetic showers. Analyzing the distributions of protons and gamma (electrons) for the ratio between a signal in the crystal containing the axis cascade and the value of the total signal in CC2 for incoming electrons and protons: $E_{MAX}^{CC2}/E_{TOT}^{CC2}$, it is possible to obtain a rejection factor ~1.6.

The information from S4 located at the bottom of the calorimeter provides a strong intrinsic rejection factor for protons, due to the difference in attenuation for hadronic and electromagnetic cascades. Electromagnetic showers initiated by gamma (electron) with initial energy up to ~100 GeV are fully contained inside a calorimeter with the thickness 16 $X_0$, while protons leave the calorimeter taking away a considerable part of the energy and produce a signal in S4. By selecting events with signals in S4 less than 40 MeV, it is possible to suppress protons with a factor of ~3.1. It turns out that a significantly more powerful criterion can be defined with distributions of the ratio of the signal in S4 to the total signal in CC2. Applying this criterion, the rejection factor ~17 is achieved.

The differences in the proton and electron cascade transverse size are also used, when analyzing information from silicon strips in CC1. The application of this criterion provides a rejection factor of ~2.

Using all presented criteria jointly, the total rejection of protons for 100-GeV electrons is ~$10^4$.

GAMMA-400 will also measure GRB spectra from lateral directions in the energy range from ~10 to ~100 MeV. The on-axis effective area is of about 0.13 m$^2$ for each of four lateral directions. The total effective field of view for the GAMMA-400 lateral aperture, when measuring GRB spectra is ~6 sr (Leonov et al., 2021).

Prototypes of some detector systems (Fig. 21) were manufactured, tested in the laboratory (Fig. 22) and calibrated on positron beams at S-25R electron synchrotron (Lebedev Physical Institute, Troitsk) in the energy range of 100-300 MeV. As a result, we obtained time resolution of ~200 ps for AC and ToF systems (Arkhangelskiy et al., 2020b) and energy resolution of 10% for CC2 at the energy of positrons of 300 MeV (Suchkov et al., 2021).

GAMMA-400 performance calculations and calibrating detectors at different accelerators will be continued to improve the performance and the event selection methods, as well as to optimize the GAMMA-400 physical scheme.

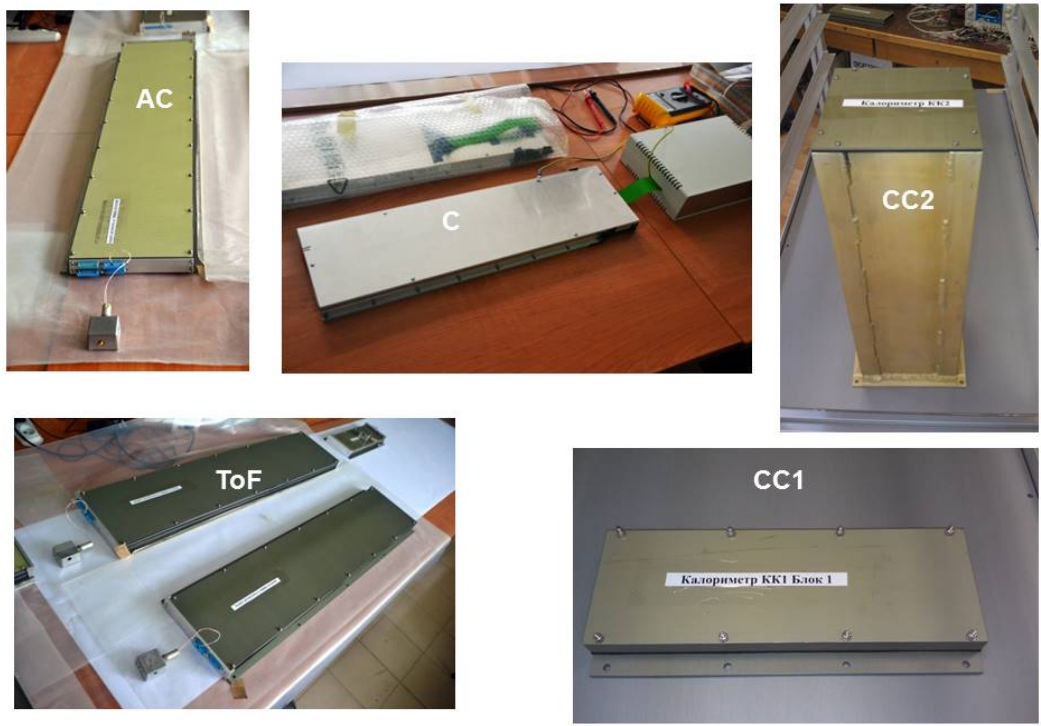

Fig. 21. Prototypes of AC, C, ToF, CC1, CC2 detector systems.

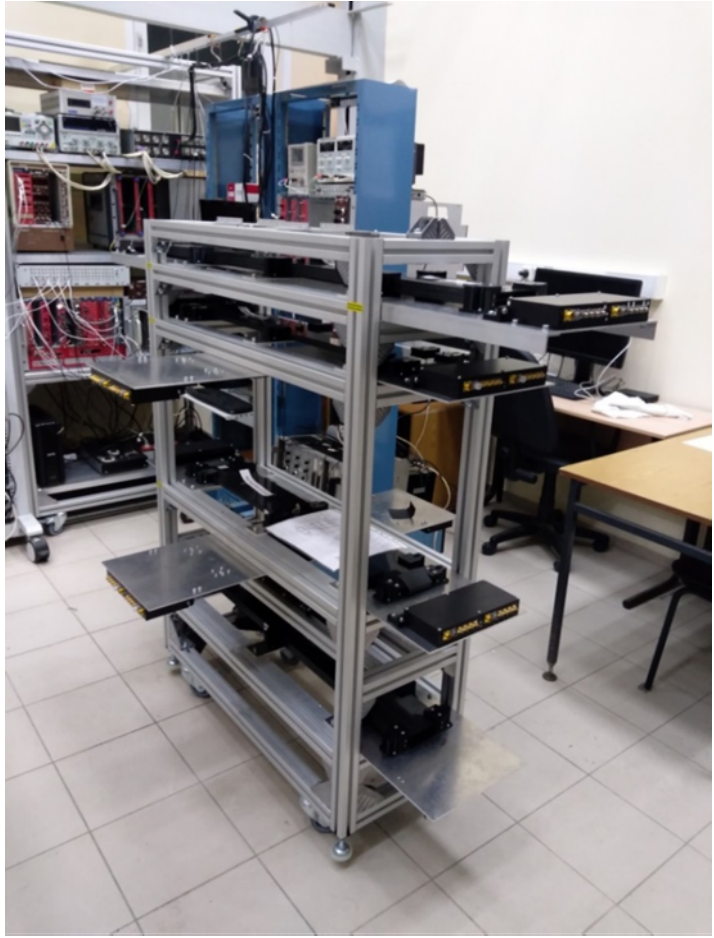

Fig. 22. Testing the GAMMA-400 detector prototypes in the laboratory.

**3.3 Comparison of GAMMA-400 with Fermi-LAT and ground-based facilities**

The GAMMA-400 gamma-ray telescope has numerous advantages in comparison with Fermi-LAT:

- highly elliptical orbit (without the Earth's occultation and away from the radiation belts) allows us to observe with the aperture of ±45° different gamma-ray sources continuously over a long period of time with 100% efficiency in contrast to 15% efficiency for Fermi-LAT operating in the scanning mode;

- thanks to use scintillating fibers and analog readout providing the best coordinate resolution in the SciFi coordinate detectors of converter-tracker and CC1, as well as time-of-flight system, GAMMA-400 has an excellent angular resolution;

- due to the deep (~18 $X_0$) calorimeter, GAMMA-400 has very good energy resolution and can provide more reliable the detection of gamma rays up to several TeV for vertically incident events;

- owing to the significant gamma-ray separation from cosmic rays (the presence of a special trigger with event timing, time-of-flight system, preshower detector S3, separation of electromagnetic and hadronic showers in shower leakage detector S4, two-layer scintillation detectors), GAMMA-400 is significantly well equipped to separate gamma rays from the background of cosmic rays and backscattering events.

GAMMA-400 will have also the unprecedented angular resolution for energy >30 GeV and very good energy resolution in the energy region 10-1000 GeV in comparison with current and future space- and ground-based instruments: VERITAS, MAGIC, H.E.S.S., CTA, and and it allows us to fill the data gap at the energy of ~100 GeV between the space- and ground-based instruments.

It is necessary to note that at present some projects of the gamma-ray telescopes are developed for further investigations of the gamma-ray emission after Fermi-LAT: e-ASTROGAM (DeAngelis et al., 2018) and AMEGO (Kierans et al., 2021) in the medium energy and HERD (Cattaneo et al., 2019) and AMS-100 (Schael et al., 2019) in high-energy range. Performance of these telescopes in comparison with GAMMA-400 and Fermi-LAT is presented in Table 2.

Table 2. Performance of space- and ground-based telescopes in comparison with GAMMA-400 and Fermi-LAT

|  | Space-based gamma-ray telescopes | | | | | | Ground-based facility |
|---|---|---|---|---|---|---|---|
|  | Medium energy | | | High-energy | | | |
|  | e-ASTROGAM | AMEGO | Fermi-LAT | GAMMA-400 | HERD | AMS-100 | CTA |
| **Country** | Europe | USA | USA/Europe | Russia | China/Europe | Europe + USA |  |
| **Energy range for gamma rays** | 0.3 MeV–3 GeV | 0.2 MeV–10 GeV | 50 MeV–1 TeV | 20 MeV–1 TeV | 0.5 GeV–10 TeV | 1 GeV–10 TeV | > 50 GeV |
| **Observation mode** | Scanning | Scanning | Scanning | Point-source | Scanning | Scanning | Scanning |
| **Orbit** | Circular, ~550 km | Circular, ~550 km | Circular, ~550 km | Highly elliptical, 500–300 000 km | Circular, ~400 km | L2 | - |
| **Angular resolution** | 0.1° ($E_\gamma$ = 1 GeV) | 1° ($E_\gamma$ = 1 GeV) | 0.1° ($E_\gamma$ = 100 GeV) | ~0.01° ($E_\gamma$ = 100 GeV) | 0.1° ($E_\gamma$ = 100 GeV) | ~0.01° ($E_\gamma$ = 100 GeV) | 0.1° ($E_\gamma$ = 100 GeV) |
| **Energy resolution** | 20% ($E_\gamma$ = 1 MeV) | 10% ($E_\gamma$ = 1 GeV) | 10% ($E_\gamma$ = 100 GeV) | ~2% ($E_\gamma$ = 100 GeV) | 1-2% ($E_\gamma$ = 100 GeV) | 1-2% ($E_\gamma$ = 100 GeV) | 15% ($E_\gamma$ = 100 GeV) |

## 4 The preliminary GAMMA-400 scientific program

The GAMMA-400 main scientific goals are: precise continuous up to ~100 days measurements of Galactic Center, Fermi Bubbles, Crab, Vela, Cygnus X, Geminga, Sun, and other regions, extended and point gamma-ray sources, GRBs, diffuse gamma rays, dark matter searching, measuring electron + positron fluxes with the angular (~0.01° at $E_\gamma$ = 100 GeV) and energy (~2% at $E_\gamma$ = 100 GeV) resolutions.

### 4.1 Investigation of gamma-ray discrete sources in the Galactic plane

Despite the fact that Fermi-LAT detected gamma-ray emission from 5065 sources, approximately 30% of gamma-ray sources (Abdollahi et al., 2020) were not identified due to insufficient angular resolution. About 40% of them are in the Galactic plane. In contrast to the Fermi-LAT scanning observation mode, GAMMA-400 will be able to identify and investigate many discrete sources due to another observation mode: long-term, continuous observations of point gamma-ray sources, as well as significantly better angular and energy resolutions.

According to the Fermi-LAT results, about 80 extended gamma-ray sources (E > 10 GeV) were found (Abdollahi et al., 2020). Thanks to long-term, continuous observations and significantly better angular and energy resolutions, GAMMA-400 will be able to clarify and map these extended sources in detail.

According to the Fermi-LAT results, an excess of gamma-ray emission from the Galactic center was detected. It is believed to be associated either with the presence of dark matter or with the presence of many millisecond pulsars. Due to long-term, continuous observations and significantly better angular and energy resolutions, GAMMA-400 will be able to clarify this problem.

GAMMA-400 will study different regions of Galactic plane, Galactic center, Fermi Bubbles, Crab, Vela, Cygnus X, Geminga and other objects with maximum FoV of ±45°.

**4.2 Dark matter searching**

Generally there are two wide classes of DM candidates, which are potentially accessible by GAMMA-400: WIMPs and ALPs.

***Galactic Center region*** historically represents the most favorable target for WIMP annihilation or decay searches due to the biggest J factor. In this context a mysterious gamma-ray excess identified a while ago (e.g. Ackermann et al., 2017) around GC attracts a great interest. It peaks at few GeV, extends up to approximately 10° around GC and can be well-fitted by annihilating WIMPs with the mass of several tens GeV and cross section around the thermal value ~$10^{-26}$ cm$^3$/s. However, alternative explanations for the origin the excess indeed exist. They include the emission from millisecond pulsars (both prompt and secondary), molecular clouds and others. The future high-resolution data from GAMMA-400 will be very important for the ultimate determination of the excess.

Another interesting aspect is the searches for hypothetical narrow spectral lines due to DM annihilation or decay directly into photons. This represents very pristine DM signature, since no other astrophysical processes are expected to produce such lines at the energies above ~0.1 GeV. We provided some estimates of GAMMA-400 sensitivity to the diphoton annihilation cross section in (Egorov et al., 2020). However our most recent simulations showed that the sensitivity presented there tends towards slightly optimistic side – realistically we would expect our sensitivity to be comparable with that of Fermi-LAT after 12-15 years of its operation. At the same time DAMPE has comparable sensitivity too (Zun-Lei Xu et al., 2021). Hence, it can be possible in the future to stack together the data from all three telescopes and significantly extend the sensitivity to narrow lines by such joint data analysis.

***ALP discovery potential by catching a nearby supernova explosion.*** Fermi-LAT has not been able to observe any supernova explosion in the Local Group due to rarity of such events. However such observation would be extremely valuable for constraining ALP properties. Since ALPs are converted into photons by the Galactic magnetic field during their propagation, gamma rays from SNe can be a signature of SN ALPs. Thus, in case of such luck, GAMMA-400 will be able to probe ALP-photon coupling constant values down to $g_{a\gamma}$ ~ $10^{-13}$ GeV$^{-1}$ for ALP masses below ~1 neV! The chance to catch such event over the mission lifetime is about 10%. Very high sensitivity of this probe still makes it one of the main GAMMA-400 objectives in the area of DM searches. The detailed calculations on this subject can be seen in (Egorov et al., 2020).

***ALP signature searches in the pulsar spectra.*** Recently the tentative ALP signature in the spectra of pulsars was identified in Fermi-LAT data (Gautham et al., 2018). However this detection is not reliable yet and the future additional GAMMA-400 observational data on those pulsars will be able potentially to confirm or deny robustly this preliminary interpretation of signal.

***Other targets*** include globular clusters, nearby galaxies, dwarf satellites, hypothesized axion clouds around neutron stars, etc.

**4.3 Studying the gamma-ray emission from the Sun**

The Sun is one of the promising targets for indirect DM searches with gamma rays. The Sun is visible in gamma rays because of the interaction of CRs with the solar environment. The standard gamma-ray emission of the Sun includes two main components: (1) the contribution from the disk, originating from the interactions of hadronic CRs with the solar atmosphere, which yield hadronic cascades with gamma-ray component, and (2) the contribution from diffuse emission, due to the interactions of cosmic-ray electrons and positrons with the optical solar photons in the heliosphere and production of gamma rays by inverse Compton scattering. The standard solar emission mechanisms are expected to yield a smooth gamma-ray spectrum, while both the DM scenarios illustrated above are expected to yield some characteristic features in the spectrum. In particular, if gamma rays are produced through a mediator, the spectrum should exhibit a boxlike feature; on the other hand, if gamma rays are produced directly in DM annihilations, a line-like feature is expected (Mazziotta et al., 2020).

The study of high-energy gamma-ray emission from solar flares is intended to elucidate the role of nuclear processes in a flare, the mechanism of particle acceleration and the interaction of accelerated beams with the solar atmosphere (Ajello et al., 2021). The GAMMA-400 gamma-ray telescope will make it possible to obtain new experimental data on high-energy gamma-ray emission from solar flares, which are necessary to refine the existing models of solar flares, namely:

1) Detailed energy spectra of gamma-ray emission during solar flares in the wide energy range from 20 MeV to several TeV;

2) Precision time profiles of the intensity of gamma-ray emission from solar flares in various energy ranges;

3) Fluxes of high-energy gamma rays from the quiet Sun.

**4.4 Searching for GRBs**

Gamma-ray bursts (GRBs) are the most violent explosions in the universe with isotropic equivalent energy release in gamma rays up to a few $10^{54}$ ergs (e.g. Minaev et al., 2020). Observations of GRBs become extremely relevant in the new era of multi-messenger astronomy, which started in 2017 since the first simultaneous detection of a binary neutron star merger GW 170817 by LIGO-Virgo gravitational wave detectors and short gamma-ray burst GRB 170817A by Fermi and INTEGRAL observatories (Abbott et al., 2017).

GRBs and their different counterparts (prompt emission, afterglow, kilonova, supernova) are successfully observed in the entire range of the electromagnetic spectrum, from radio to high-energy gamma rays. However, a unified model describing all observational features has not been proposed yet. This is largely due to the extremely difficult implementation of simultaneous observations in different ranges of the energy spectrum, especially at the active phase stage (prompt emission), which lasts typically only a few seconds.

Gamma rays with the energy $E_\gamma$ > 20 MeV were first detected in the EGRET/CGRO experiment for 28 gamma-ray bursts in the period 1991 - 2000 (Kaneko et al., 2008). At present, the high-energy component of gamma-ray bursts is recording mainly in the Fermi-LAT experiment, which recorded so far about 200 gamma-ray bursts in the range above 30 MeV (Ajello et al., 2019).

In most GRBs the duration of high-energy emission is significantly (up to tens of times) greater than the duration of low-energy emission and usually has a power-law like decaying light curve. It is also observed as an additional component in the energy spectrum, described by a power-law model (Ajello et al., 2019). It indicates the high-energy component in such GRBs represents an additional component to regular gamma-ray emission for these cases. Nevertheless, in some GRBs, the temporal profile of the burst at high energies does indeed repeat the profile at low energies, which implies the same nature of the emission and the absence of an additional component at high energies. Thus, a bimodal behavior of gamma-ray bursts at high energies is presented. The nature of the additional high-energy component and the conditions for its appearance has not been clarified, which makes its study an actual and important task.

An open question is the maximum possible energy of photons emitted in gamma-ray bursts. Recently sub-TeV high-energy emission was confirmed for GRB 180720B by H.E.S.S. in (0.1, 0.44) TeV energy range (Abdalla et al., 2019) and for GRB 190114C by MAGIC in (0.1, 1) TeV energy range (Acciari et al., 2019).

Using top-down and lateral directions of GAMMA-400 will allow us to detect new GRBs with rate of 10-15 GRBs per year to solve problems of GRB science. The most prominent results will be obtained in case of simultaneous GRB registration in a wide energy range of electromagnetic channel (from radio waves to high-energy gamma rays) by different ground-based and space-based facilities, as well as in gravitational wave channel.

**4.5 Studying CR electron + positron fluxes**

High-energy cosmic-ray electrons provide a unique probe of nearby to the Solar System cosmic sources. Because electrons in the interstellar space fast lose energy via inverse Compton scattering and synchrotron emission their diffusion length is about 1 kpc at energy above 1 TeV. It means that potential sources are located in the vicinity of the Solar System (Kobayashi et al., 2004). A precise measurement of the electron + positron spectrum at high energies above several hundreds of GeV might reveal interesting spectral features and flux anisotropy due to a specific source. Moreover, the anomaly increase of the positron fraction in electron and positron fluxes over 10 GeV established by PAMELA (Adriani et al., 2009) and AMS-02 (Accardo et al., 2014) experiments can require a primary source of the positrons in addition to the commonly accepted secondary origin of the positrons.

Astrophysical objects like pulsars and Super Novae Remnants (SNR) or exotic objects like clumps of hypothetical dark matter are more plausible and investigated candidates for such sources. The electrons and positrons spectrum would exhibit a different spectral features due to origin of the primary source of electrons and positrons, in the energy range above 10 GeV and, especially at about 1 TeV, possible mass of hypothetical DM particles.

It is highly desirable to measure spectrum of electrons + positrons in the TeV region with much more statistics accuracy than it is possible now. Moreover, the data of direct measurements (DAMPE, CALET) in the TeV region show a suppression of the flux (DAMPE Collaboration, 2017, Adriani et al. 2018). GAMMA-400 with high acceptance both in the main (~18$X_0$) and lateral apertures (~43$X_0$) (see Table 1) will be able to obtain accurate spectral shape above 1 TeV with energy resolution ~2% (which will decrease up to ~5% at 10 TeV due to leakage from the calorimeter) and, possible, to extend accurate measurements to 20-30 TeV using the lateral aperture, needed to understand origin of local sources.

**5 Conclusions**

The future space-based GAMMA-400 gamma-ray telescope will operate onboard the Russian astrophysical observatory in a highly elliptic orbit during 7 years to observe Galactic plane, Galactic Center, Fermi Bubbles, Crab, Vela, Cygnus X, Geminga, Sun, and other regions and measure gamma- and cosmic-ray fluxes. Observations will be performed in the point-source mode continuously for a long time (~100 days) in contrast to the scanning mode for Fermi-LAT. GAMMA-400 will have the unprecedented angular resolution for energies >30 GeV (~0.01° at $E_\gamma$ = 100 GeV) better than the Fermi-LAT and ground-based gamma-ray telescopes by a factor of 5-10, as well as very good energy resolution (~2% at $E_\gamma$ = 100 GeV). Significant separation of gamma rays from cosmic-ray background, as well as electrons + positrons from protons will allow us to measure gamma rays in the energy range from ~20 MeV to several TeV and cosmic-ray electrons + positrons up to several tens TeV. GAMMA-400 observations will permit to resolve gamma rays from annihilation or decay of dark matter particles, identify many discrete sources, clarify the structure of extended sources, specify the data on cosmic-ray electron + positron spectra.

After Fermi-LAT the GAMMA-400 mission will greatly improve the direct data on low-energy and high-energy gamma-ray and electron + positron fluxes due to unprecedented angular and very good energy resolutions, large area, and point-source continuous long-term simultaneous coaxial gamma-ray and X-ray telescope observations. The launch of the GAMMA-400 space observatory is scheduled for ~2030.

**Funding**

This study was funded by the Russian State Space Corporation ROSCOSMOS and in part by the Ministry of Science and Higher Education of the Russian Federation under Project "Fundamental problems of cosmic rays and dark matter" (contract no. 0723-2020-0040).

**References**

Abbott, B. P., Abbott, R., Abbott, T. D. et al., 2017. Multi-messenger Observations of a Binary Neutron Star Merger. Astrophys. J. Lett., 848, L12.

Abdalla, H., Adam, R., Aharonian, F. et al., 2019. A very-high-energy component deep in the gamma-ray burst afterglow. Nature 575, 7783, 464-467.

Abdollahi,S., Ackermann, M., Ajello, M/, et al., 2017, Cosmic-ray electron-positron spectrum from 7 GeV to 2 TeV with the Fermi Large Area Telescope. Phys. Rev. D 95, 082007.

Abdollahi, S., Acero, F., Ackermann, M., Ajello, M., et al., 2020. Fermi large area telescope fourth source catalog. Astrophys. J. Suppl. Ser. 247, 33, 1-37.

Accardo, L., Aguilar, M., Aisa, D., et al., 2014. High statistics measurement of the positron fraction in primary cosmic rays of 0.5–500 GeV with the Alpha Magnetic Spectrometer on the International Space Station. Phys. Rev. Lett. 113, 121101, 1-9.

Acciari, V. A., Ansoldi, S., Antonelli, L. A. et al., 2019. Teraelectronvolt emission from the γ-ray burst GRB 190114C. Nature 575, 7783, 455-458.

Ackermann, M., Ajello, M., Albert, A., et al., 2017. The Fermi galactic center GeV excess and implications for dark matter. Astrophys. J. 840, 43, 1-34.

Adriani, O., Barbarino, G., Bazilevskaya, G., et al., 2009. An anomalous positron abundance in cosmic rays with energies 1.5–100 GeV. Nature 458, 607-609.

Adriani, O., Barbarino, G., Bazilevskaya, G., et al., 2014. The PAMELA mission: heralding a new era in precision cosmic ray physics. Phys. Reports 544, 323-370.

Adriani, O., Barbarino, G., Bazilevskaya, G., et al., 2017. Ten years of PAMELA in space. Nuevo Cimento 10, 473-522.

Adriani, O., Akaike, Y., Asano, K., et al., 2018. Extended measurement of the cosmic-ray electron and positron spectrum from 11 GeV to 4.8 TeV with the calorimetric electron telescope on the International Space Station. Phys. Rev. Lett. 120, 261102.

Aguilar, M, Aisa, D., Alpat, B., et al., 2014. Precision Measurement of the ($e^+$+$e^-$) Flux in Primary Cosmic Rays from 0.5 GeV to 1 TeV with the Alpha Magnetic Spectrometer on the International Space Station. . Phys. Rev. Lett. 113, 221102.

Aguilar, M, Alberti, G., Alpat, B., et al., 2013. First result from the Alpha Magnetic Spectrometer on the International Space Station: precision measurement of the positron fraction in primary cosmic rays of 0.5–350 GeV. Phys. Rev. Lett. 110, 141102, 1-10.

Aharonian, F., Akhperjanian, A., Anton, G., et al., 2009. Probing the ATIC peak in the cosmic-ray electron spectrum with H.E.S.S. Astron. Astrophys. 508, 561-564.

Ajello, M., Arimoto, M., Axelsson, M., et al., 2019. A decade of gamma-ray bursts observed by Fermi-LAT: The second GRB catalog. Astrophys. J. 878, 52, 1-61.

Ajello, M., Baldini, L., Bastieri, D., et al., 2021. First Fermi-LAT Solar Flare Catalog. Astrophys. J. SS 252, 13, 1-31.

Ambrosi, G., An, Q., Asfandiyarov, R., et al. Direct detection of a break in the teraelectronvolt cosmic-ray spectrum of electrons and positrons. Nature 552, 63-66.

Arkhangelskiy, A., Galper, A., Arkhangelskaja, I., et al., 2020a. Design of the readout electronics for the fast trigger and time of flight of the GAMMA-400 gamma-ray telescope. J. Phys. Conf. Ser. 1690, 012024, 1-5.

Arkhangelskiy, A., Galper, A., Arkhangelskaja, I., et al., 2020b. The anticoincidence system of space-based gamma-ray telescope GAMMA-400, test beam studies of anticoincidence detector prototype with SiPM readout. Phys. At. Nucl. 83, 252-257.

Atwood, W., Abdo, A., Ackermann, M., et al., 2009. The large area telescope on the Fermi gamma-ray telescope mission. Astrophys. J. 697, 1071-1102.

Calore, F., Carenza, P., Giannotti, M., et al., 2020. Bounds on axionlike particles from the diffuse supernova flux. Phys. Rev. D 102, 123005, 1-12.

Cattaneo, P.W. on behalf of the HERD collaboration, 2019. The space station based detector HERD: precise high energy cosmic rays physics and multimessenger astronomy. Nucl. and Particle Phys. Proc. 306-308, 85-91.

Galper, A., Adriani, O., Aptekar, A., et al., 2013. Status of the GAMMA-400 project. Adv. Space Res. 51, 297-300.
Galper, A., Topchiev, N., and Yurkin, Yu., 2018. GAMMA-400 project. Astron. Rep. 62, 882-889.
Gautham, A., Calore, F., Carenza, P., et al., 2020. Reconciling hints on axion-like-particles from high-energy gamma rays with stellar bounds. arXiv:2008.08100.
Chang, J., Adams, J., Ahn, H., et al., 2008. An excess of cosmic ray electrons at energies of 300–800 GeV. Nature 456, 362-365.
Chang, J., Ambrosi, G., An, Q., et al., 2017. The DArk Matter Particle Explorer mission. Astropart. Phys. 95, 6-24.
DAMPE Collaboration, 2017. Direct detection of a break in the teraelectronvolt cosmic-ray spectrum of electrons and positrons. Nature 552, 63–66.
DeAngelis, A., Tatischeff, V., Grenier, I., et al., 2018. Science with e-ASTROGAM. A space mission for MeV–GeV gamma-ray astrophysics. J. High Energy Astroph. 19, 1-106.
Egorov, A., Topchiev, N., Galper, A., et al., 2020. Dark matter searches by the planned gamma-ray telescope GAMMA-400. J. Cosmol. Astropart. Phys. 11, 049, 1-25.
Evoli, C., Amato, E., Blasi, P., and Aloisio, R., 2021. Galactic factories of cosmic-ray electrons and positrons. Phys. Rev. D 103, 083010, 1-22.
Kaneko, Y., Magdalena González, M., Preece, R., et al., 2008. Broadband spectral properties of bright high-energy gamma-ray bursts observed with BATSE and EGRET. Astrophys. J. 677, 1168-1183.
Kardashev, N., Khartov, V., Abramov, V., et al., 2013. "RadioAstron"—A Telescope with a Size of 300 000 km: Main Parameters and First Observational Results. Astron. Rep. 57, 3, 153–194.
Kierans, C., Harding, A., Cenko, B., et al., 2021. AMEGO: exploring the extreme multimessenger universe. arXiv:2101.03105.
Kirn, T., on behalf of the LHCb collaboration, 2017. SciFi – A largescintillating fibretrackerforLHCb. Nucl. Instr. Meth. Phys. Res. A 845, 481-485.
Kobayashi, T., Komori, Y., Yoshida, K., and Nishimura, J., 2004. The most likely sources of high-energy cosmic-ray electrons in supernova remnants. Astrophys. J. 601, 340-351.
Leane, R., Slatyer, T., Beacom, J., et al., 2018. GeV-scale thermal WIMPs: Not even slightly ruled out. Phys. Rev. D 98, 023016, 1-16.
Leonov, A., Galper, A., Bonvicini, V., et al., 2015. Separation of electrons and protons in the GAMMA-400 gamma-ray telescope. Adv. Space Res. 56, 1538-1545.
Leonov, A., Galper, A., Topchiev, N., et al., 2019. Capabilities of the GAMMA-400 gamma-ray telescope for observation of electrons and positrons in the TeV energy range. Phys. At. Nucl. 82, 855-858.
Leonov, A., Galper, A., Topchiev, N., et al., 2021. Capabilities of the GAMMA-400 gamma-ray telescope to detect gamma-ray bursts from lateral directions. Adv. Space Res., https://doi.org/10.1016/j.asr.2021.10.031; arXiv:2103.07161.
Lien, A., Sakamoto, T., Barthelmy, S. et al., 2016. The Third SWIFT Burst Alert Telescope Gamma-Ray Burst Catalog. Astrophys. J. 829, 7.
Mazziotta, M., Loparco, F., Serini, D., et al., 2020. Search for dark matter signatures in the gamma-ray emission towards the Sun with the Fermi Large Area Telescope. Phys. Rev. D 102, 022003, 1-12.
Minaev, P., and Pozanenko, A., 2020. The $E_{p,i}$-$E_{iso}$ correlation: type I gamma-ray bursts and the new classification method. MNRAS 492, 1919-1936.
Mori, N., on behalf of the CALET collaboration, 2013. CALET: a calorimeter for cosmic-ray measurements in space. Nucl. Phys. B (Proc. Suppl.) 239-240, 199-203.
Schael, S., Atanasyan, A., J. Berdugo, J., et al., 2019. AMS-100: The next generation magnetic spectrometer in space – An international science platform for physics and astrophysics at Lagrange point 2. Nuclear Inst. and Methods in Physics Research A 944, 162561, 1-13.
Staszak, D. for the VERITAS Collaboration, 2015. A Cosmic-ray Electron Spectrum with VERITAS, arXiv:1508.06597.

Suchkov, S., Galper, A., Arkhangelskiy, A., et al., 2021. Calibrating the prototype calorimeter for the GAMMA-400 gamma-ray telescope on the positron beam at the Pakhra accelerator. Instrum. Exp. Tech. 64, 669-675.

Topchiev, N., Galper, A., Bonvicini, V., et al., 2017. High-energy gamma-ray studying with GAMMA-400 after Fermi-LAT. J. Phys. Conf. Ser., 798, 012011, 1-6.

Topchiev, N., Galper, A., Arkhangelskaja, I., et al., 2019. The future space-based GAMMA-400 gamma-ray telescope for studying gamma and cosmic rays. Bull. Russ. Acad. Sci. Phys. 83, 629-631.

Tridon, D., Colin, P., Cossio, L., et al., 2011. Measurement of the cosmic electron plus positron spectrum with the MAGIC telescopes. arXiv:1110.4008.

von Kienlin, A., Meegan, C., Paciesas, et al., 2020. The Fourth Fermi-GBM Gamma-Ray Burst Catalog: A Decade of Data. Astrophys. J. 893, 46.

Zun-Lei Xu, Kai-Kai Duan, Wei Jiang, et al., 2021. Optimal gamma-ray selections for monochromatic line searches with DAMPE. arXiv:2107.13208.